\documentclass[11pt,a4paper]{article}
\pdfoutput=1
\usepackage{jheppub}
\usepackage{tikz}
\usetikzlibrary{intersections, calc, arrows.meta}
\usepackage{subcaption}
\usepackage{bm}

\DeclareMathOperator{\Tr}{Tr}

\newcommand{\ri}{\mathrm{i}}

\newcommand{\cob}{\delta}

\newcommand{\ve}{\varepsilon}

\newcommand{\hf}{\frac{1}{2}}
\newcommand{\qu}{\frac{1}{4}}
\newcommand{\til}[1]{\widetilde{#1}}
\newcommand{\si}{\sigma}

\renewcommand{\b}[1]{\overline{#1}}
\newcommand{\del}{\partial}

\newcommand{\lap}{\Delta}
\newcommand{\bra}{\langle}
\newcommand{\ket}{\rangle}
\newcommand{\la}{\lambda}

\newcommand{\ka}{\kappa}

\newcommand{\bt}{\beta}
\newcommand{\ga}{\gamma}

\newcommand{\al}{\alpha}

\newcommand{\rt}[1]{\sqrt{#1}}
\newcommand{\cO}{\mathcal{O}}
\newcommand{\cZ}{\mathcal{Z}}

\newcommand{\cM}{\mathcal{M}}

\newcommand{\cI}{\mathcal{I}}

\newcommand{\bbZ}{{\mathbb Z}}
\newcommand{\gs}{g_{\rm s}}
\newcommand{\conn}{{\rm c}}
\newcommand{\bbra}[1]{\langle{#1}|}
\newcommand{\kket}[1]{|{#1}\rangle}

\begin{document}

\title{FZZT branes in JT gravity and topological gravity}

\author[a]{Kazumi Okuyama}
\author[b]{and Kazuhiro Sakai}

\affiliation[a]{Department of Physics, Shinshu University,\\
3-1-1 Asahi, Matsumoto 390-8621, Japan}
\affiliation[b]{Institute of Physics, Meiji Gakuin University,\\
1518 Kamikurata-cho, Totsuka-ku, Yokohama 244-8539, Japan}

\emailAdd{kazumi@azusa.shinshu-u.ac.jp, kzhrsakai@gmail.com}

\abstract{
We study Fateev-Zamolodchikov-Zamolodchikov-Teschner (FZZT)
branes in Witten-Kontsevich topological gravity,
which includes Jackiw-Teitelboim (JT) gravity as a special case.
Adding FZZT branes to topological gravity
corresponds to inserting determinant operators
in the dual matrix integral and amounts to a certain shift of
the infinitely many couplings of topological gravity.
We clarify the perturbative interpretation of adding FZZT branes
in the genus expansion of topological gravity in terms of
a simple boundary factor and the generalized Weil-Petersson volumes.
As a concrete illustration we study JT gravity in the presence of
FZZT branes and discuss its relation to
the deformations of the dilaton potential
that give rise to conical defects.
We then construct a non-perturbative formulation of FZZT branes
and derive a closed expression for the general correlation
function of multiple FZZT branes and multiple macroscopic loops.
As an application we study the FZZT-macroscopic loop correlators
in the Airy case. We observe numerically a void in the eigenvalue
density due to the eigenvalue repulsion induced by FZZT-branes
and also the oscillatory behavior
of the spectral form factor which is expected
from the picture of eigenbranes.}

\maketitle

\section{Introduction}
Two-dimensional Jackiw-Teitelboim (JT) gravity \cite{Jackiw:1984je,Teitelboim:1983ux} is a useful toy model
to study various aspects of quantum gravity and holography.
In a remarkable paper \cite{Saad:2019lba} it was shown that
JT gravity is holographically dual to a certain double-scaled matrix model.
This is an example of the holography involving the ensemble average, where
the Hamiltonian $H$ of the boundary theory becomes the random
matrix.
This in particular implies that the correlator of two partition functions
$\bra Z(\bt_1)Z(\bt_2)\ket$ includes the contribution of spacetime wormhole
connecting the two asymptotic boundaries and thus the correlator is not factorized.

As argued in \cite{Blommaert:2019wfy},
one can fix some of the eigenvalues of the matrix integral
by introducing the 
Fateev-Zamolodchikov-Zamolodchikov-Teschner (FZZT)
branes \cite{Fateev:2000ik,Teschner:2000md},
which are called ``eigenbranes'' in \cite{Blommaert:2019wfy}.
Adding branes in JT gravity is also considered in \cite{Penington:2019kki}
to recover the Page curve for black hole evaporation.
In a recent paper \cite{Gao:2021uro}, the matrix model
for the dynamical end of the world (EOW) brane in JT gravity is proposed.\footnote{See also \cite{Goel:2020yxl} for a classification of branes in JT gravity.}

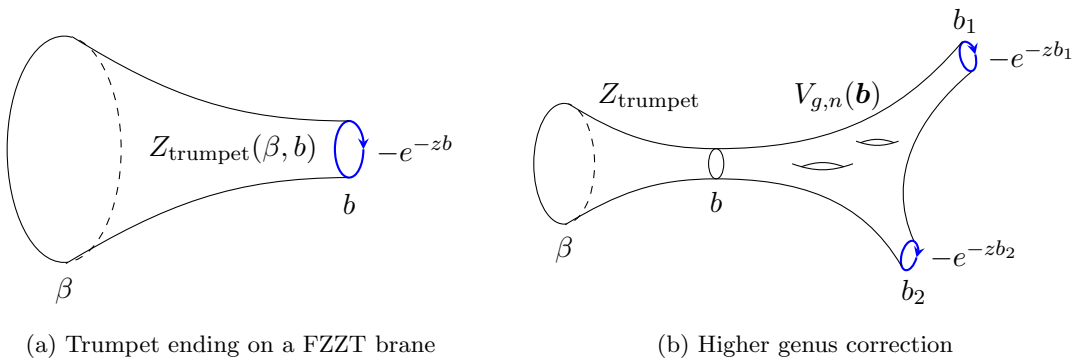
\begin{figure}[h]
\captionsetup[subfigure]{font=footnotesize}
\centering
\subcaptionbox{Trumpet ending on a FZZT brane \label{sfig:trumpet-M}}[.5\textwidth]{%
\begin{tikzpicture}[scale=0.75]
\draw (0,2) arc [start angle=90,end angle=270, x radius=1, y radius=2];
\draw[dashed] (0,2) arc [start angle=90,end angle=-90, x radius=1, y radius=2];
\draw[->,>=stealth, thick, blue] (5.25,0) arc [start angle=360,end angle=0, x radius=0.25, y radius=0.5];
\draw (0.14,1.99) to [out=-32,in=180] (5,0.5);
\draw (0.03,-2.02) to [out=31,in=180] (5,-0.5);
\draw (0,-2.1) node [below]{$\bt$};
\draw (5,-0.6) node [below]{$b$};
\draw (5.3,0) node [right]{$-e^{-zb}$};
\draw (1.3,0) node [right]{$Z_{\text{trumpet}}(\bt,b)$};
\end{tikzpicture}
}%
\centering
\subcaptionbox{Higher genus correction \label{sfig:trumpet-V}}[.5\textwidth]{%
\begin{tikzpicture}[scale=0.4]
\draw (0,2) arc [start angle=90,end angle=270, x radius=1, y radius=2];
\draw[dashed] (0,2) arc [start angle=90,end angle=-90, x radius=1, y radius=2];
\draw (5,0) circle [x radius=0.25, y radius=0.5];
\draw (0.14,1.99) to [out=-32,in=180] (5,0.5);
\draw (0.03,-2.02) to [out=31,in=180] (5,-0.5);
\draw (0,-2.1) node [below]{$\bt$};
\draw (5,-0.6) node [below]{$b$};
\draw[->,>=stealth, thick, blue,rotate=-15] (12,0) arc [start angle=360,end angle=0, x radius=0.25, y radius=0.5];
\draw[->,>=stealth, thick, blue,rotate=15] (14,0) arc [start angle=360,end angle=0, x radius=0.25, y radius=0.5];
\draw (5,0.5) to [out=0,in=-130] (13.1,4.05);
\draw (5,-0.5) to [out=0,in=120] (11.1,-3.4);
\draw (13.4,3.05) to [out=-140,in=114] (11.5,-2.55);
\draw (7.5,0) to [out=-20,in=200] (9.5,0);
\draw (7.8,-0.1) to [out=20,in=160] (9.2,-0.1);
\draw (9.6,0.75) to [out=-20,in=200] (11,0.75);
\draw (9.75,0.7) to [out=20,in=160] (10.85,0.7);
\draw (13.2,4) node [above]{$b_1$};
\draw (11.5,-3.5) node [below]{$b_2$};
\draw (13.6,3.6) node [right]{$-e^{-zb_1}$};
\draw (11.7,-3) node [right]{$-e^{-zb_2}$};
\draw (9,1.4) node [above]{$V_{g,n}(\bm{b})$};
\draw (2.8,1.4) node [above]{$Z_{\text{trumpet}}$};
\end{tikzpicture}
}%
\caption{
Trumpet can end on a FZZT brane. We attach the factor
$\cM(b)=-e^{-zb}$ to the geodesic boundary
and integrate over $b$.
}
\label{fig:trumpet-FZZT}
\end{figure}

\begin{figure}[h]
\captionsetup[subfigure]{font=footnotesize}
\centering
\subcaptionbox{Wormhole \label{sfig:wormhole}}[.5\textwidth]{%
\begin{tikzpicture}[scale=0.4]
\draw (0,2) arc [start angle=90,end angle=270, x radius=1, y radius=2];
\draw[dashed] (0,2) arc [start angle=90,end angle=-90, x radius=1, y radius=2];
\draw (5,0.5) arc [start angle=90,end angle=270, x radius=0.25, y radius=0.5];
\draw[dashed] (5,0.5) arc [start angle=90,end angle=-90, x radius=0.25, y radius=0.5];
\draw (0.14,1.99) to [out=-32,in=180] (5,0.5);
\draw (0.03,-2.02) to [out=31,in=180] (5,-0.5);
\draw (0,-2.1) node [below]{$\bt_1$};
\draw (5,-0.6) node [below]{$b$};
\draw (10,0) circle [x radius=1, y radius=2];
\draw (5,0.5) to [out=0,in=212] (9.8,1.97);
\draw (5,-0.5) to [out=0,in=149] (9.8,-1.97);
\draw (10,-2.1) node [below]{$\bt_2$};
\end{tikzpicture}
}%
\centering
\subcaptionbox{Half-wormholes \label{sfig:half-wormhole}}[.5\textwidth]{%
\begin{tikzpicture}[scale=0.4]
\draw (0,2) arc [start angle=90,end angle=270, x radius=1, y radius=2];
\draw[dashed] (0,2) arc [start angle=90,end angle=-90, x radius=1, y radius=2];
\draw[->,>=stealth, thick, blue] (5.25,0) arc [start angle=360,end angle=0, x radius=0.25, y radius=0.5];
\draw (0.14,1.99) to [out=-32,in=180] (5,0.5);
\draw (0.03,-2.02) to [out=31,in=180] (5,-0.5);
\draw (0,-2.1) node [below]{$\bt_1$};
\draw (10.8,0) circle [x radius=1, y radius=2];
\draw (5.7,0.5) to [out=0,in=212] (10.6,1.97);
\draw (5.7,-0.5) to [out=0,in=149] (10.6,-1.97);
\draw (10.8,-2.1) node [below]{$\bt_2$};
\draw[thick, blue] (5.7,0.5) arc [start angle=90,end angle=270, x radius=0.25, y radius=0.5];
\draw[dashed, thick, blue] (5.7,0.5) arc [start angle=90,end angle=-90, x radius=0.25, y radius=0.5];
\draw[->,>=stealth, thick, blue] (5.45,-0.01)--(5.45,0.01);
\draw (4.9,-0.6) node [below]{$z$};
\draw (5.8,-0.6) node [below]{$z$};
\end{tikzpicture}
}%
\caption{Contributions of  \subref{sfig:wormhole} wormhole and \subref{sfig:half-wormhole} half-wormholes.
}
  \label{fig:half-wormhole}
\end{figure}
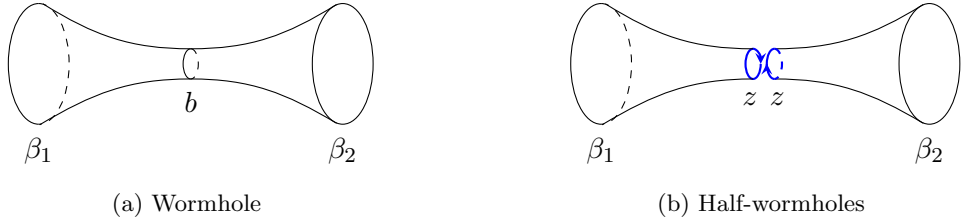

In this paper, we consider FZZT branes in JT gravity.
Our prescription for introducing the FZZT branes
is a natural generalization of the JT gravity matrix model
by Saad, Shenker and Stanford \cite{Saad:2019lba}.
We obtain the amplitude in the presence of FZZT branes by gluing several 
building blocks. As shown in \cite{Saad:2019lba}, 
the JT gravity amplitude is obtained by gluing the ``trumpet'' \eqref{eq:def-trumpet}
and the Weil-Petersson (WP) volume \eqref{eq:V-WP}.
A new ingredient in the presence of a FZZT brane is the factor
$\cM(b)=-e^{-zb}$ with $z$ being a parameter, which is attached to the geodesic boundary of
length $b$ and we integrate over $b$ (see Figure~\ref{fig:trumpet-FZZT}).
This construction can be generalized to multiple FZZT branes by
replacing the factor with $\cM(b)=-\sum_i e^{-z_ib}$.
In particular, the trumpet can end on a FZZT brane as shown in
Figure~\ref{sfig:trumpet-M}.
This implies that the two-boundary correlator
$\bra Z(\bt_1)Z(\bt_2)\ket$ in the presence of FZZT brane receives
a contribution depicted in Figure~\ref{sfig:half-wormhole}, which
reminds us of the ``half-wormhole'' introduced in \cite{Saad:2021rcu}. 

It turns out that the above construction of FZZT branes in JT gravity
can be generalized to encompass
arbitrary background $\{t_k\}$ of 2d topological gravity.
We can compute FZZT brane amplitude by gluing certain building blocks.
In general background $\{t_k\}$, we can use the same ``trumpet'' and $\cM(b)$
as in JT gravity, but the WP volume should be replaced by the generalized WP volume
defined in \eqref{eq:Vgn-operator}.

This construction defines an FZZT brane amplitude only perturbatively 
in genus expansion.
One can obtain the non-perturbative expression of 
the correlator of FZZT branes and macroscopic loop operators $Z(\bt)=\Tr e^{-\bt H}$
by taking a double scaling limit of the
correlator of determinants in the finite $N$ matrix model.
We find a compact expression of the generating function of
the $Z(\bt)$-FZZT correlators in terms of the Baker-Akhiezer (BA) function
and the Christoffel-Darboux (CD) kernel (see \eqref{eq:even-gen} and \eqref{eq:odd-gen}).
As an application of our formalism, we consider the
spectral form factor in the presence of two FZZT branes (see Figure~\ref{fig:SFF}).

This paper is organized as follows. 
In section~\ref{sec:genus},
we find the prescription for the construction of
FZZT brane amplitude by gluing the trumpet, (generalized) WP volume and $\cM(b)$.
We find that the half-wormhole amplitude is given by the complementary error function
\eqref{eq:z-HW}.
We also comment on the relation between the trumpet and the Liouville wavefunction.
In section~\ref{sec:deform},
we compute the genus-zero density of states in JT gravity with
$K$ FZZT branes in the 't Hooft limit \eqref{eq:thooft}.
We also comment on the relation to the deformation of the potential of
dilaton gravity.
In section~\ref{sec:BA},
we review the known results
about the BA function, the CD kernel and the multi-FZZT amplitude.
In section~\ref{sec:corr},
we find the generating functions 
\eqref{eq:even-gen} and \eqref{eq:odd-gen} of the correlators of 
FZZT branes and macroscopic loop operators.
In section~\ref{sec:airy},
we apply our formalism to the Airy case, i.e.~the trivial background
$t_k=0~(k\geq1)$ of topological gravity.
We study the
spectral form factor in the presence of two FZZT branes
in the Airy case.
Finally we conclude in section~\ref{sec:conclusion}.
In appendix~\ref{app:minimal} we summarize the result of $(2,p)$ minimal string.
In appendix~\ref{app:inverse} we provide an alternative
derivation of \eqref{eq:Z-FZZT-full} using the correlator of inverse determinant.
\section{FZZT branes in the genus expansion}\label{sec:genus}
In this section, we will explain our formalism of
computing the genus expansion of the
correlator of partition functions
$Z(\bt)=\Tr e^{-\bt H}$ in the presence of FZZT branes.

\subsection{JT gravity and Weil-Petersson volume}
Let us first consider JT gravity.
As shown by Saad, Shenker and Stanford \cite{Saad:2019lba},
the genus expansion of the connected correlator of $Z(\bt)$'s can be obtained by gluing
the ``trumpet'' and the Weil-Petersson (WP) volume
\begin{equation}
\begin{aligned}
 \left\bra\prod_{i=1}^n Z(\bt_i)\right\ket_\conn
=\sum_{g=0}^\infty e^{-(2g-2+n)S_0}
\prod_{i=1}^n \int_0^\infty \tilde{b}_id\tilde{b}_i 
\tilde{Z}_{\text{trumpet}}(\bt,\tilde{b}_i)\tilde{V}_{g,n}(\tilde{b}_1,\ldots,\tilde{b}_n),
\end{aligned} 
\label{eq:corrZ-orig}
\end{equation} 
where the subscript $\conn$ of $\bra\cdots\ket_\conn$ refers to the connected part.
The WP volume $\tilde{V}_{g,n}(\bm{\tilde{b}})$ is given by
\begin{equation}
\begin{aligned}
 \tilde{V}_{g,n}(\bm{\tilde{b}})&=\int_{\b{\cM}_{g,n}}e^{2\pi^2\ka+
\hf\sum_{i=1}^n \tilde{b}_i^2\psi_i},
\end{aligned} 
\label{eq:WPdef}
\end{equation}
where $\overline{\cal M}_{g,n}$
denotes the Deligne-Mumford compactification of the moduli space of
${\cal M}_{g,n}$ of Riemann surface $\Sigma$ of genus $g$
with $n$ marked points $p_1,\ldots,p_n$,
$\kappa$
is the first Miller-Morita-Mumford
class and
$\psi_i$ is the first Chern class of the complex line bundle
whose fiber is the cotangent space to $p_i$.
Note that \eqref{eq:WPdef} is valid for $3g-3+n\ge 0$
and $\tilde{V}_{0,1}, \tilde{V}_{0,2}$ are undefined.
Correspondingly, \eqref{eq:corrZ-orig} makes sense
except the contributions of the disk and annulus amplitudes,
which are to be discussed separately.
The trumpet partition function $\tilde{Z}_{\text{trumpet}}(\bt,\tilde{b})$
in \eqref{eq:corrZ-orig} is given by
\begin{equation}
\begin{aligned}
 \tilde{Z}_{\text{trumpet}}(\bt,\tilde{b})=\frac{\rt{\ga}}{\rt{2\pi\bt}}e^{-\frac{\ga\tilde{b}^2}{2\bt}},
\end{aligned} 
\end{equation}
where $\ga$ is the asymptotic value of the dilaton field
near the boundary of $AdS_2$.

It is convenient to set
\begin{equation}
\begin{aligned}
 \ga=\frac{1}{2\pi^2},\quad \tilde{b}^2=2\pi^2 b^2.
\end{aligned} 
\end{equation}
Then we find
\begin{equation}
\begin{aligned}
 \tilde{V}_{g,n}(\bm{\tilde{b}})&=
\int_{\b{\cM}_{g,n}}e^{2\pi^2(\ka+
\hf\sum_i b_i^2\psi_i)}=(2\pi^2)^{3g-3+n}V_{g,n}(\bm{b}),
\end{aligned} 
\end{equation}
where we defined the rescaled WP volume by
\begin{equation}
\begin{aligned}
 V_{g,n}(\bm{b})=\int_{\b{\cM}_{g,n}}e^{\ka+
\hf\sum_{i=1}^n b_i^2\psi_i}.
\end{aligned} 
\label{eq:V-WP}
\end{equation}
In what follows we will call $V_{g,n}(\bm{b})$ the WP volume instead of the original
$\tilde{V}_{g,n}(\bm{\tilde{b}})$. As in our previous paper
\cite{Okuyama:2019xbv} we define the genus counting parameter $\gs$ by
\begin{equation}
\begin{aligned}
 \gs=(2\pi^2)^{\frac{3}{2}}e^{-S_0}.
\end{aligned} 
\end{equation}
It is also convenient to rescale the trumpet so that
\begin{equation}
\begin{aligned}
e^{-S_0(2g-2+n)}\tilde{V}_{g,n}(\bm{\tilde{b}})
\prod_{i=1}^n\tilde{b}_id\tilde{b}_i\tilde{Z}_{\text{trumpet}}(\bt,\tilde{b}_i)
=\gs^{2g-2+n}V_{g,n}(\bm{b})\prod_{i=1}^n b_idb_i 
Z_{\text{trumpet}}(\bt,b_i).
\end{aligned} 
\end{equation}
Then we find
\begin{equation}
\begin{aligned}
Z_{\text{trumpet}}(\bt,b)=\frac{e^{-\frac{b^2}{2\bt}}}{\rt{2\pi\bt}}.
\end{aligned} 
\label{eq:def-trumpet}
\end{equation}
In this new normalization, 
\eqref{eq:corrZ-orig} is written as
\begin{equation}
\begin{aligned}
 \left\bra\prod_{i=1}^n Z(\bt_i)\right\ket_\conn
=\sum_{g=0}^\infty \gs^{2g-2+n}
\prod_{i=1}^n \int_0^\infty b_idb_i
Z_{\text{trumpet}}(\bt,b_i)V_{g,n}(\bm{b}).
\end{aligned} 
\label{eq:corrZ}
\end{equation}

Now let us consider the effect of adding FZZT branes. In the matrix model the FZZT brane
corresponds to the insertion of the
determinant operator $\det(\xi+H)$ in the matrix integral,
where $H$ is the matrix variable and $\xi$ is a formal parameter
\cite{Kutasov:2004fg,Maldacena:2004sn}.\footnote{\label{foot:M}In
the literature of the matrix model of 2d gravity, the macroscopic loop
is defined by $Z(\bt)=\Tr e^{\bt M}$. On the other hand,
in JT gravity $Z(\bt)=\Tr e^{-\bt H}$ is interpreted as the partition function
of boundary theory. Thus the random matrix $M$ and the Hamiltonian $H$ are related by
\begin{equation}
\begin{aligned}
 M=-H.
\end{aligned} 
\end{equation}
This implies that the FZZT brane operator
\eqref{eq:det-fermion} is written as $\det(\xi-M)$.
}
The FZZT brane operator $\det(\xi+H)$
can be represented by means of
the vector degrees of freedom as
\begin{equation}
\begin{aligned}
 \det(\xi+H)=\int  d\chi d\b{\chi}\, e^{\b{\chi}(\xi+H)\chi},
\end{aligned} 
\label{eq:det-fermion}
\end{equation}
where $\chi$ and  $\b{\chi}$ are Grassmann-odd vector variables.

From the relation
\begin{equation}
\begin{aligned}
 \det(\xi+H)=e^{\Tr\log(\xi+H)}=\sum_{n=0}^\infty\frac{1}{n!}\Bigl[
\Tr\log(\xi+H)\Bigr]^n,
\end{aligned} 
\end{equation}
we see that the insertion of FZZT brane creates infinitely many boundaries
corresponding to the single trace operator $\Tr\log(\xi+H)$.
This single trace operator has an integral representation
\begin{equation}
\begin{aligned}
-\int_{\ve}^\infty \frac{d\bt}{\bt}e^{-\xi\bt}Z(\bt)&= -\int_{\ve}^\infty \frac{d\bt}{\bt}e^{-\xi\bt}\Tr e^{-\bt H}\\
&=
\Tr\log(\xi+H)+\log\ve+\cO(\ve^0),
\end{aligned} 
\label{eq:trlog-int}
\end{equation}
where $\ve$ is a small regularization parameter.
The logarithmically divergent term $\log\ve$ can be absorbed into
the overall normalization of the determinant operator and 
we will ignore such divergence unless otherwise stated.

We can apply the integral transformation \eqref{eq:trlog-int} to
some of the $Z(\bt)$'s in \eqref{eq:corrZ}.
It turns out that the integral transformation of the trumpet partition
function is free from the logarithmic divergence. We introduce
$\cM(b)$ by\footnote{In \cite{Gao:2021uro}, this integral transformation is called
the ``inverse trumpet''.}
\begin{equation}
\begin{aligned}
 \frac{1}{b}\cM(b)=-\int_0^\infty\frac{d\bt}{\bt}e^{-\xi\bt}Z_{\text{trumpet}}(\bt,b)
=-\int_0^\infty\frac{d\bt}{\bt}e^{-\xi\bt}\frac{e^{-\frac{b^2}{2\bt}}}{\rt{2\pi\bt}}.
\end{aligned} 
\label{eq:cM-int}
\end{equation}
Using the integral representation of the modified Bessel function
$K_{\hf}(x)$
\begin{equation}
\begin{aligned}
 K_{\hf}(x)=\hf\rt{\frac{x}{2}}\int_0^\infty dt\, t^{-\frac{3}{2}}
e^{-t-\frac{x^2}{4t}}=\rt{\frac{\pi}{2x}}e^{-x},
\end{aligned} 
\label{eq:K-hf}
\end{equation}
we find that $\cM(b)$ in \eqref{eq:cM-int} is given by
\begin{equation}
\begin{aligned}
 \cM(b)=-e^{-zb},
\end{aligned} 
\label{eq:cMb}
\end{equation}
where $\xi$ and $z$ are related by
\begin{equation}
\begin{aligned}
 \xi=\hf z^2.
\end{aligned} 
\label{eq:xi-z}
\end{equation}
Note that the integral representation \eqref{eq:K-hf} is valid
when $|\text{arg}(x)|<\pi/4$, so that
the integral \eqref{eq:cM-int} is equal to
\eqref{eq:cMb} under the condition
\begin{equation}
\begin{aligned}
 -\frac{\pi}{4}<\text{arg}(z)<\frac{\pi}{4}.
\end{aligned} 
\label{eq:arg-z}
\end{equation}
This in particular implies that the expression of $\cM(b)$ in \eqref{eq:cMb}
is valid when $\text{Re}(\xi)=\text{Re}(z^2/2)>0$.

Thus the connected correlator of $Z(\bt)$'s in the presence of FZZT brane
is given by
\begin{equation}
\begin{aligned}
 &\left\bra\det(\xi+H)\prod_{i=1}^m Z(\bt_i)\right\ket_\conn\\
&=\sum_{n=0}^\infty
\frac{1}{n!}\left\bra\Bigl[\Tr\log(\xi+H)\Bigr]^n\prod_{i=1}^m Z(\bt_i)\right\ket_\conn\\
&=\sum_{n=0}^\infty\frac{(-1)^n}{n!}
\prod_{j=1}^n\int_0^\infty\frac{d\bt_j'}{\bt_j'}e^{-\xi\bt_j'}
\left\bra \prod_{j=1}^n Z(\bt_j')\prod_{i=1}^m Z(\bt_i)\right\ket_\conn\\
&=\sum_{n=0}^\infty\frac{(-1)^n}{n!}
\prod_{j=1}^n\int_0^\infty\frac{d\bt_j'}{\bt_j'}e^{-\xi\bt_j'}\sum_{g=0}^\infty
\gs^{2g-2+n+m}\\
&\times\prod_{j=1}^n\int_0^\infty b_j'db_j'Z_{\text{trumpet}}(\bt_j',b_j')
\prod_{i=1}^m\int_0^\infty b_idb_iZ_{\text{trumpet}}(\bt_i,b_i)V_{g,n+m}(\bm{b'},\bm{b})\\
&=\sum_{g,n=0}^\infty\frac{\gs^{2g-2+n+m}}{n!}
\prod_{j=1}^n\int_0^\infty db_j'\cM(b_j')
\prod_{i=1}^m\int_0^\infty b_idb_iZ_{\text{trumpet}}(\bt_i,b_i)V_{g,n+m}(\bm{b'},\bm{b}),
\end{aligned} 
\label{eq:FZZT-formula}
\end{equation}
where we used the definition of $\cM(b)$ in \eqref{eq:cM-int}.
This expression is valid except 
the terms of the order of $\gs^k\ (k=-1,0)$,
which can be calculated separately.
To summarize, we can introduce the FZZT branes in the
correlator of $Z(\bt)$'s by gluing $\cM(b)$ along the geodesic boundary
and integrate over $b$ with measure $\cM(b)db$ (see Figure~\ref{fig:trumpet-FZZT}).
Note that the factor of $b$ in the integration measure $bdb$
for the trumpet is canceled out
by the $1/b$ in \eqref{eq:cM-int}.

We can interpret our expression of \eqref{eq:cMb} as follows.
$\cM(b)$ in \eqref{eq:cMb} has the form
\begin{equation}
\begin{aligned}
 \cM(b)=-e^{-S_{\text{particle}}}
\end{aligned} 
\label{eq:cM-particle}
\end{equation}
where the particle action $S_{\text{particle}}$ is given by
\begin{equation}
\begin{aligned}
 S_{\text{particle}}=zb=(\text{mass})\times (\text{length of worldline}).
\end{aligned} 
\end{equation}
Namely, $\cM(b)$ can be interpreted as the contribution of
a particle with mass $z$ running around the geodesic boundary with length $b$.
The overall minus sign in \eqref{eq:cM-particle} comes from the fact
that the particle running around the loop is a fermion (see \eqref{eq:det-fermion}).

If we change the sign of $\cM(b)$
\begin{equation}
\begin{aligned}
 \cM(b)=+e^{-zb},
\end{aligned} 
\end{equation} 
it corresponds to the anti-FZZT brane
represented by the inverse determinant
\begin{equation}
\begin{aligned}
 \det(\xi+H)^{-1}=\int d\phi d\b{\phi} \,e^{\b{\phi}(\xi+H)\phi}
\end{aligned} 
\end{equation}
where $\phi$ and $\b{\phi}$ are the Grassmann-even (bosonic) vector degrees of freedom.

We can generalize our formula \eqref{eq:FZZT-formula} to include
multiple FZZT branes.
Using the relation
\begin{equation}
\begin{aligned}
 \prod_i\det(\xi_i+H)=\exp\left[\sum_i \Tr(\xi_i+H)\right]
=\exp\left[-\sum_i\int_0^\infty\frac{d\bt}{\bt}e^{-\xi_i\bt}Z(\bt)\right]
\end{aligned} 
\end{equation}
we can use the same formula \eqref{eq:FZZT-formula}
for the correlator of $Z(\bt)$'s in the presence of multiple FZZT branes
by simply replacing
\begin{equation}
\begin{aligned}
 \cM(b)=-\sum_i e^{-z_ib},
\end{aligned} 
\label{eq:multi-cM}
\end{equation}
where $z_i$ is related to $\xi_i$ by $\xi_i=z_i^2/2$.

In \cite{Gao:2021uro}, the matrix model description of the end of the world (EOW)
brane
in JT gravity is considered. The prescription of \cite{Gao:2021uro}
is the same as our formula
\eqref{eq:FZZT-formula} with $\cM(b)$ replaced by
\begin{equation}
\begin{aligned}
 \cM_{\text{EOW}}(b)=\frac{e^{-\mu b}}{2\sinh\frac{b}{2}}=\sum_{n=0}^\infty
e^{-(\mu+n+\hf)b}.
\end{aligned} 
\label{eq:EOW}
\end{equation}
In our interpretation, this corresponds to infinitely many anti-FZZT branes
with a particular set of parameters $z_n=\mu+n+\hf~(n\geq0)$.
As discussed in \cite{Gao:2021uro}, the integral of \eqref{eq:EOW} has a divergence
coming from the pole at $b=0$ and  a certain regularization is required
to define the EOW brane. On the other hand, in our case of FZZT brane $\cM(b)$
in \eqref{eq:cMb}
has no pole at $b=0$ and the $b$-integral is well-defined. 

\subsection{FZZT branes in general background of topological gravity}
It turns out that the above construction can be generalized
to encompass arbitrary background $\{t_k\}$ of 2d topological gravity.
Recall that JT gravity is a special case of topological gravity
with infinitely many couplings turned on as
$t_k=\ga_k$ where \cite{Mulase:2006baa,Dijkgraaf:2018vnm,Okuyama:2019xbv}
\begin{equation}
\begin{aligned}
 \ga_0=\ga_1=0,\quad \ga_k=\frac{(-1)^k}{(k-1)!}\quad(k\geq2).
\end{aligned} 
\label{eq:ga-JT}
\end{equation}
In Witten-Kontsevich topological gravity
\cite{Witten:1990hr,Kontsevich:1992ti}  observables
are made up of the intersection numbers
\begin{align}
\langle\tau_{d_1}\cdots\tau_{d_n}\rangle_{g,n}
 =\int_{\overline{\cal M}_{g,n}}
  \psi_1^{d_1}\cdots\psi_n^{d_n},\qquad d_1,\ldots,d_n\in\mathbb{Z}_{\ge 0}.
\label{eq:intsec}
\end{align}
The generating function
for the intersection numbers is defined as
\begin{align}
F(\{t_k\})
 :=\sum_{g=0}^\infty \gs^{2g-2}F_g(\{t_k\}),\qquad
F_g(\{t_k\})
 :=\left\langle e^{\sum_{d=0}^\infty t_d\tau_d}\right\rangle_g.
\label{eq:genF}
\end{align}

We assume that the trumpet is background independent
(i.e.~independent of $t_k$)
\begin{equation}
\begin{aligned}
 Z_{\text{trumpet}}(\bt,b)=\frac{e^{-\frac{b^2}{2\bt}}}{\rt{2\pi\bt}}.
\end{aligned} 
\end{equation}
In section~\ref{sec:Liouville} we will consider the validity of this assumption.
Then it is natural to define the generalized WP volume $V_{g,n}$ by
\begin{equation}
\begin{aligned}
 \bra Z(\bt_1)\cdots Z(\bt_n)\ket_\conn =:\sum_{g=0}^\infty\gs^{2g-2+n}
\prod_{i=1}^n\int_0^\infty b_idb_i
Z_{\text{trumpet}}(\bt_i,b_i)V_{g,n}(\bm{b},\{t_k\}).
\end{aligned} 
\label{eq:gen-V}
\end{equation}
Note that the correlator of $Z(\bt)$'s can be written as
\begin{equation}
\begin{aligned}
 \bra Z(\bt_1)\cdots Z(\bt_n)\ket_\conn=B(\bt_1)\cdots B(\bt_n)F(\{t_k\})
\end{aligned} 
\label{eq:ZcorrB}
\end{equation}
where $B(\bt)$ is the boundary creation operator \cite{Moore:1991ir}
\begin{equation}
\begin{aligned}
 B(\bt)=\frac{\gs}{\rt{2\pi}}\sum_{k=0}^\infty\bt^{k+\hf}\del_k.
\end{aligned} 
\end{equation}
As in the previous section
\eqref{eq:gen-V} and \eqref{eq:ZcorrB} are valid except
the $(g,n)=(0,1),(0,2)$ parts and we leave $V_{0,1},V_{0,2}$ undefined.
Here we are interested in the general structure and do not go into
the details of these parts.
See e.g.~\cite{Okuyama:2020ncd} for a precise treatment of them.

Let us introduce the operator $V(b)$ by
\begin{equation}
\begin{aligned}
 B(\bt) =:\gs\int_0^\infty bdb
Z_{\text{trumpet}}(\bt,b)V(b).
\end{aligned} 
\label{eq:BtoV}
\end{equation}
Then we find
\begin{equation}
\begin{aligned}
 V(b)=\sum_{k=0}^\infty \frac{b^{2k}}{2^kk!}\del_k.
\end{aligned} 
\end{equation}
From \eqref{eq:gen-V}, \eqref{eq:ZcorrB} and \eqref{eq:BtoV} we find
that the generalized WP volume has a simple expression
\begin{equation}
\begin{aligned}
 V_{g,n}(\bm{b},\{t_k\})=V(b_1)\cdots V(b_n)F_g(\{t_k\}).
\end{aligned} 
\label{eq:Vgn-operator}
\end{equation}
From \eqref{eq:genF}, this is also written as
\begin{equation}
\begin{aligned}
 V_{g,n}(\bm{b},\{t_k\})&=\sum_{k_1,\ldots,k_n=0}^\infty
\left\bra \exp\left(\sum_{d=0}^\infty t_d\tau_d\right)\prod_{i=1}^n \frac{b_i^{2k_i}}{2^{k_i}k_i!}
\psi_i^{k_i}\right\ket_g\\
&=\left\bra \exp\left(\sum_{d=0}^\infty t_d\tau_d
+\hf\sum_{i=1}^n b_i^2\psi_i\right)\right\ket_g.
\end{aligned} 
\label{eq:gen-Vbis}
\end{equation}
One can see that this is a natural generalization of the WP volume \eqref{eq:V-WP}.
With our definition of the generalized WP volume in \eqref{eq:gen-V}, 
the computation of the correlator of $Z(\bt)$'s in the presence of FZZT branes
is completely parallel to that in the previous subsection for
the JT gravity case \eqref{eq:FZZT-formula}.
Thus we conclude that the FZZT brane amplitude of topological gravity 
in the general background $\{t_k\}$ is given by the same formula
\eqref{eq:FZZT-formula} with the WP volume replaced by the generalized WP volume
in \eqref{eq:gen-V}.

As an application of our formalism \eqref{eq:FZZT-formula}, let us consider the
FZZT brane amplitude without $Z(\bt)$-insertions
\begin{equation}
\begin{aligned}
 \left\bra\prod_{i}\det(\xi_i+H)\right\ket_\conn&=\sum_{g,n=0}^\infty
\frac{\gs^{2g-2+n}}{n!}\prod_{j=1}^n\int_0^\infty db_j\cM(b_j)V_{g,n}(\bm{b},\{t_k\})
\end{aligned} 
\end{equation}
where $\cM(b)$ is given by \eqref{eq:multi-cM}.
Since $V_{0,1},V_{0,2}$ are undefined, the calculation here
is valid for positive powers of $\gs$.
From the relation
\begin{equation}
\begin{aligned}
 \int_0^\infty db\cM(b)V(b)=-\sum_{k=0}^\infty\sum_i (2k-1)!!z_i^{-2k-1}\del_k
\end{aligned} 
\end{equation}
we find
\begin{equation}
\begin{aligned}
 \left\bra\prod_{i}\det(\xi_i+H)\right\ket_\conn&=\sum_{n=0}^\infty 
\frac{1}{n!}\left[-\sum_{k=0}^\infty\sum_i\gs(2k-1)!!z_i^{-2k-1}\del_k\right]^nF(\{t_k\})\\
&=\exp\left(-\sum_{k=0}^\infty\sum_i\gs(2k-1)!!z_i^{-2k-1}\del_k\right)F(\{t_k\})\\
&=F(\{\til{t}_k\}),
\end{aligned} 
\end{equation}
where 
\begin{equation}
\begin{aligned}
 \til{t}_k=t_k-\gs(2k-1)!!\sum_i z_i^{-2k-1}.
\end{aligned} 
\label{eq:tilde-tk}
\end{equation}
This shift of coupling $t_k$
due to the insertion of the FZZT branes
agrees with the known result in the literature\footnote{In
the literature the convention
$x_{2k+1}\equiv \dfrac{t_k}{\gs(2k+1)!!}$ is often used
for the couplings.
In terms of $x_\ell$ the shift \eqref{eq:tilde-tk} is written as
\begin{align}
\til{x}_\ell=x_\ell-\sum_i\frac{1}{\ell}\frac{1}{z_i^\ell}
\qquad (\ell:\mbox{odd}).
\end{align}}
\cite{Date:1982yeu} (see e.g.~\cite{dubrovin}
for the expression in our convention).
In the theory of soliton equations
the shift is generated by the action of
the vertex operator of \cite{Lepowsky:1978jk},
which is viewed as the infinitesimal B\"acklund transformation
for the KdV equation \cite{Date:1981qx}.
When $t_k=0$,
this expression of the couplings in terms of $z_i$
\eqref{eq:tilde-tk} has appeared in the Kontsevich
matrix model \cite{Kontsevich:1992ti} and
it is known as the Miwa transform.
In the context of minimal string theory \cite{Seiberg:2003nm,Seiberg:2004at},
this shift is interpreted as the open-closed duality where
the insertion of FZZT branes is replaced by the shift of closed string
background $\{t_k\}$ \cite{Maldacena:2004sn,Gaiotto:2003yb}.

\subsection{Example of generalized WP volume}
Let us compute a few examples of the generalized WP volume $V_{g,n}(\bm{b})$
for small $g$ and $n$.
One can compute $V_{g,n}(\bm{b})$
from the known result of the correlator of $Z(\bt)$'s
using the relation \eqref{eq:gen-V}.

The generalized WP volume at genus-zero
has been computed in \cite{Mertens:2020hbs}.
From the result of the genus-zero $n$-point function of $Z(\bt)$
\cite{Moore:1991ir}
\begin{equation}
\begin{aligned}
 \bra Z(\bt_1)\cdots Z(\bt_n)\ket_\conn^{g=0}=\gs^{n-2}\del_0^{n-3}\left[\del_0u_0
\prod_{i=1}\rt{\frac{\bt_i}{2\pi}}e^{\bt_i u_0}\right],
\end{aligned} 
\label{eq:Zcorr-g0}
\end{equation}
one can read off $V_{0,n}(\bm{b})$ using \eqref{eq:gen-V}.
Here $u_0=\del_0^2F_0$.
To this end, it is convenient to define the function
$\cI_k(b)=\cI_k(b;u_0)$ by
\begin{equation}
\begin{aligned}
 \int_0^\infty bdb
 Z_{\text{trumpet}}(\bt,b)\cI_k(b)=\frac{\bt^{k+\hf}}{\rt{2\pi}}e^{\bt u_0}
 \qquad(k\in\bbZ_{\ge 0}).
\end{aligned} 
\label{eq:trumpet-Iint}
\end{equation}
One can easily show that $\cI_k(b)$ is given by
\begin{equation}
\begin{aligned}
 \cI_k(b)&=\left(\frac{b}{\rt{2u_0}}\right)^kI_k(b\rt{2u_0})\\
&=\sum_{n=0}^\infty\frac{u_0^n\left({b^2}/{2}\right)^{n+k}}{n!(n+k)!},
\end{aligned} 
\label{eq:cI-def}
\end{equation}
where $I_k(x)$ denotes the modified Bessel function of the first kind.
The following property of $\cI_k(b)$ is also useful
\begin{equation}
\begin{aligned}
 \int_0^\infty db e^{-zb}\cI_k(b)=\frac{(2k-1)!!}{(z^2-2u_0)^{k+\hf}}.
\end{aligned} 
\label{eq:cM-Ik-int}
\end{equation}
From \eqref{eq:Zcorr-g0} we find that the generalized WP volume at genus-zero is given by
\begin{equation}
\begin{aligned}
 V_{0,n}(\bm{b})=\del_0^{n-3}\left[\del_0u_0\prod_{i=1}^n\cI_0(b_i)\right].
\end{aligned}
\label{eq:V0n} 
\end{equation}
This agrees with the result of \cite{Mertens:2020hbs}.

At genus-one, from our previous result of the correlators
of $Z(\bt)$'s \cite{Okuyama:2020ncd,Okuyama:2021ytf} 
\begin{equation}
\begin{aligned}
\bra Z(\bt)\ket^{g=1}
&=\gs\rt{\frac{\bt}{2\pi}}e^{\beta u_0}\Biggl(
\frac{I_2}{24t^2}+\frac{\beta}{24t}\Biggr),\\
\bra Z(\beta_1)Z(\beta_2)\ket_\conn^{g=1}
&=\gs^2\frac{\sqrt{\beta_1\beta_2}}{2\pi}e^{(\beta_1+\beta_2)u_0}
 \Biggl(
  \frac{I_3}{24t^3}
 +\frac{I_2^2}{12t^4}
 +\frac{I_2(\beta_1+\beta_2)}{12t^3}
 +\frac{\beta_1^2+\beta_1\beta_2+\beta_2^2}{24t^2}
\Biggr),
\end{aligned} 
\end{equation}
we find that $V_{1,1}(b)$ and $V_{1,2}(b_1,b_2)$ are given by
\begin{equation}
\begin{aligned}
 V_{1,1}(b)&=\frac{I_2\cI_0(b)}{24t^2}+\frac{\cI_1(b)}{24t},\\
V_{1,2}(b_1,b_2)&=
 \frac{I_3\cI_0(b_1)\cI_0(b_2)}{24t^3}
 +\frac{I_2^2\cI_0(b_1)\cI_0(b_2)}{12t^4}
+\frac{I_2(\cI_1(b_1)\cI_0(b_2)+\cI_0(b_1)\cI_1(b_2))}{12t^3}\\
& +\frac{\cI_2(b_1)\cI_0(b_2)+\cI_1(b_1)\cI_1(b_2)+\cI_0(b_1)\cI_2(b_2)}{24t^2}.
\end{aligned} 
\label{eq:Vgn-example}
\end{equation}
Here $I_k$ is the Itzykson-Zuber variable \cite{Itzykson:1992ya}
\begin{equation}
\begin{aligned}
 I_k=\sum_{n=0}^\infty t_{k+n}\frac{u_0^n}{n!},
\qquad t=1-I_1.
\end{aligned} 
\label{eq:IZvar}
\end{equation}

Alternatively, we can compute $V_{g,n}(\bm{b})$ using \eqref{eq:Vgn-operator}.
Let us compute $V_{1,1}(b)$ in this way. Applying
the operator $V(b)$ to
the genus-one free energy $F_1=-\frac{1}{24}\log t$ we find
\begin{equation}
\begin{aligned}
 V_{1,1}(b)&=V(b)F_1=\frac{1}{24t}V(b)I_1.
\end{aligned} 
\end{equation}
Using the relation
\begin{equation}
\begin{aligned}
 \del_k I_n=\frac{I_{n+1}}{t}\frac{u_0^k}{k!}+\frac{u_0^{k-n}}{(k-n)!}
 \theta[k-n],\qquad \theta[n]:=\left\{\begin{array}{ll}
    0\quad&n<0,\\ 1&n\ge 0,\end{array}\right.
\end{aligned}
\end{equation}
we find
\begin{equation}
\begin{aligned}
 V_{1,1}(b)&=\frac{1}{24t}\sum_{k=0}^\infty \frac{b^{2k}}{2^kk!}\left[
\frac{I_{2}}{t}\frac{u_0^k}{k!}+\frac{u_0^{k-1}}{(k-1)!}\theta[k-1]\right]\\
&=\frac{I_2\cI_0(b)}{24t^2}+\frac{\cI_1(b)}{24t}.
\end{aligned} 
\end{equation}
This agrees with the result in \eqref{eq:Vgn-example} obtained from
$\bra Z(\bt)\ket^{g=1}$.

We should stress that the generalized WP volume $V_{g,n}(\bm{b})$ in not a polynomial in $b_i$ for general background $\{t_k\}$. However, when $u_0=0$ $V_{g,n}(\bm{b})$
becomes a polynomial since
\begin{equation}
\begin{aligned}
 \lim_{u_0\to0}\cI_k(b)=\frac{b^{2k}}{2^kk!}.
\end{aligned} 
\end{equation}
For instance, when $u_0=0$ $V_{1,1}(b)$ becomes
\begin{equation}
\begin{aligned}
 \lim_{u_0\to0}V_{1,1}(b)=\frac{t_2}{24(1-t_1)^2}+\frac{b^2}{48(1-t_1)}.
\end{aligned} 
\end{equation}
In particular, 
in the JT gravity case $t_k=\ga_k$ \eqref{eq:ga-JT}
we have $u_0=0$.
One can check that the generalized WP volume reduces to the WP
volume when $t_k=\ga_k$.

A certain generalization of the WP volume is considered in \cite{Mertens:2020hbs}
in the minimal model background. Our definition of $V_{g,n}(\bm{b})$ is different
from \cite{Mertens:2020hbs}.

\subsection{Half-wormholes and factorization}

In this section we consider the genus-zero amplitude
between $Z(\bt)$ and FZZT brane, which we call the ``half-wormhole''
following \cite{Saad:2021rcu}.\footnote{See also \cite{Mukhametzhanov:2021nea,Choudhury:2021nal,Garcia-Garcia:2021squ,Saad:2021uzi}.}
The half-wormhole amplitude is easily found as
\begin{equation}
\begin{aligned}
 Z_{\text{HW}}(\bt,z)&=\int_0^\infty db Z_{\text{trumpet}}(\bt,b)\cM(b)\\
&=\int_0^\infty db Z_{\text{trumpet}}(\bt,b)(-e^{-zb})\\
&=-\hf e^{\hf z^2\bt}\text{Erfc}\left(\rt{\frac{z^2\bt}{2}}\right),
\end{aligned} 
\label{eq:z-HW}
\end{equation}
where $\text{Erfc}(x)$ denotes the complementary error function
\begin{equation}
\begin{aligned}
 \text{Erfc}(x)=\frac{2}{\rt{\pi}}\int_{x}^\infty dt e^{-t^2}.
\end{aligned} 
\end{equation}
On the other hand, the ``wormhole'' amplitude is obtained by gluing two trumpets
\begin{equation}
\begin{aligned}
 \bra Z(\bt_1)Z(\bt_2)\ket_\conn^{g=0}&=\int_0^\infty bdb Z_{\text{trumpet}}(\bt_1,b)
Z_{\text{trumpet}}(\bt_2,b)=\frac{\rt{\bt_1\bt_2}}{2\pi(\bt_1+\bt_2)},
\end{aligned} 
\end{equation}
where we assumed $u_0=0$.
In the presence of FZZT brane, the spacetime of JT gravity can end on 
the FZZT brane along the geodesic boundary. 
For instance, let us consider the amplitude with two asymptotic boundaries
$Z(\bt_1)$ and $Z(\bt_2)$ with one FZZT brane $\det(\xi+H)$
\begin{equation}
\begin{aligned}
 \bra Z(\bt_1)Z(\bt_2)\det(\xi+H)\ket.
\end{aligned} 
\label{eq:ZZ-FZZT}
\end{equation}
At the order of $\cO(\gs^0)$,
there are two contributions as shown in Figure~\ref{fig:half-wormhole}.
The contribution of Figure~\ref{sfig:half-wormhole} is factorized
\begin{equation}
\begin{aligned}
Z_{\text{HW}}(\bt_1,z)Z_{\text{HW}}(\bt_2,z). 
\end{aligned} 
\end{equation}
Of course, the total amplitude \eqref{eq:ZZ-FZZT} is not factorized. But
if we increase the number of FZZT branes as in \eqref{eq:multi-cM}
the trumpet can end on various branes labeled by $z_i$ and
there are many contributions like Figure~\ref{sfig:half-wormhole}
with various choices of $z_i$. 
This is a concrete realization of the idea of ``eigenbrane'' introduced
in \cite{Blommaert:2019wfy}.
As discussed in \cite{Blommaert:2019wfy}, adding FZZT branes
corresponds to fixing some of the eigenvalues of matrix.\footnote{Fixing the eigenvalue 
of the random matrix model is also discussed in
\cite{Blommaert:2021gha}.}
This suggests that adding many FZZT branes amounts to pick a particular
member of the ensemble of random matrices and we expect 
that the factorization is restored, at least partially.

\subsection{Relation to Liouville wavefunction}\label{sec:Liouville}
In the traditional approach to 2d gravity, the correlator of 
macroscopic loop operators is commonly written in terms of 
the (mini-superspace)
wavefunction of the Liouville theory \cite{Moore:1991ir,Ginsparg:1993is}
\begin{equation}
\begin{aligned}
 \psi_E(\bt)=\frac{1}{\pi}\rt{2E\sinh\pi E}\,K_{\ri E}(\bt E_0)
\end{aligned} 
\label{eq:psi-E}
\end{equation}
which satisfies
\begin{equation}
\begin{aligned}
 \Bigl[-(\bt\del_\bt)^2+E_0^2\bt^2\Bigr]\psi_E(\bt)=E^2\psi_E(\bt).
\end{aligned} 
\end{equation}
In \eqref{eq:psi-E} $K_\nu(x)$ denotes the modified Bessel function of the second kind
and we defined
\begin{equation}
\begin{aligned}
 E_0=-u_0.
\end{aligned} 
\end{equation}
The wavefunction $\psi_E(\bt)$ in \eqref{eq:psi-E}
is normalized as\footnote{Our normalization
of $\psi_E(\ell)$ is different from that in \cite{Moore:1991ir,Ginsparg:1993is}
by a factor of $\rt{2}$. To correctly normalize it as \eqref{eq:psi-delta}
we need the extra factor of $\rt{2}$.}
\begin{equation}
\begin{aligned}
 \int_0^\infty\frac{d\bt}{\bt}\psi_E(\bt)\psi_{E'}(\bt)&=\cob(E-E'),\\
\int_0^\infty dE\psi_E(\bt)\psi_{E}(\bt')&=\bt\cob(\bt-\bt').
\end{aligned} 
\label{eq:psi-delta}
\end{equation}

In our formalism \eqref{eq:FZZT-formula}, the correlator of macroscopic loop operators
is obtained by gluing the trumpet and the generalized WP volume.
In the traditional approach to 2d gravity,
the trumpet has not appeared in the literature before, as far as we know.
The trumpet partition function naturally appears in JT gravity
from the path integral of Schwarzian mode describing the wiggles
near the asymptotic boundary of $AdS_2$ \cite{Saad:2019lba}.
One might think that the trumpet is tightly connected to JT gravity
and it cannot be generalized to the topological gravity with arbitrary
background $\{t_k\}$.
However, as we will see below it turns out that the trumpet partition function
is written in terms of the Liouville wavefunction $\psi_E(\bt)$,
which connects our formalism \eqref{eq:FZZT-formula} and the traditional approach.
Thus we can use the trumpet in arbitrary
background $\{t_k\}$ as we did in the previous section.

In order to write $Z_{\text{trumpet}}(\bt,b)$ in terms of
 $\psi_E(\bt)$, we start with
the relation \eqref{eq:cM-int}
\begin{equation}
\begin{aligned}
 \int_0^\infty\frac{d\bt}{\bt}e^{-\hf z^2\bt}Z_{\text{trumpet}}(\bt,b)
=\frac{e^{-zb}}{b}.
\end{aligned} 
\label{eq:trumpet-Mb}
\end{equation}
It is convenient to set
\begin{equation}
\begin{aligned}
 z=2\rt{E_0}\cosh\frac{\si}{2}.
\end{aligned} 
\label{eq:z-si}
\end{equation}
Then \eqref{eq:trumpet-Mb} becomes
\begin{equation}
\begin{aligned}
 \int_0^\infty\frac{d\bt}{\bt}e^{-\bt E_0\cosh\si-\bt E_0}Z_{\text{trumpet}}(\bt,b)
=\frac{1}{b}e^{-2\rt{E_0}b\cosh\frac{\si}{2}}.
\end{aligned} 
\label{eq:trumpet-cosh}
\end{equation}
Multiplying
both sides of \eqref{eq:trumpet-cosh} by $\cos(E\si)$
and using the relation \cite{Martinec:2003ka}
\begin{equation}
\begin{aligned}
 \int_0^\infty d\si e^{-\bt E_0\cosh\si}\cos(E\si)=K_{\ri E}(\bt E_0),
\end{aligned} 
\end{equation}
we find
\begin{equation}
\begin{aligned}
 \int_0^\infty\frac{d\bt}{\bt}K_{\ri E}(\bt E_0)e^{-\bt E_0}Z_{\text{trumpet}}(\bt,b)
&=\int_0^\infty d\si\cos(E\si)\frac{e^{-2\rt{E_0}b\cosh\frac{\si}{2}}}{b}\\
&=\frac{2}{b}K_{2\ri E}(2b\rt{E_0}).
\end{aligned} 
\end{equation}
By using the orthogonality of $\psi_E(\bt)$ in \eqref{eq:psi-delta},
this is inverted as
\begin{equation}
\begin{aligned}
e^{-\bt E_0} Z_{\text{trumpet}}(\bt,b)&=\frac{4}{\pi^2 b}
\int_0^\infty dE E\sinh\pi E\,K_{\ri E}(\bt E_0)K_{2\ri E}(2b\rt{E_0})\\
&=\frac{1}{b}\int_0^\infty dE\frac{\psi_E(\bt)\psi_{2E}(2b/\rt{E_0})}{\rt{\cosh\pi E}}.
\end{aligned} 
\label{eq:trumpet-wave}
\end{equation}
This is our desired result: the trumpet partition function can be expressed
in terms of the Liouville wavefunction $\psi_E(\bt)$.

Let us consider the wormhole amplitude
\begin{equation}
\begin{aligned}
 \bra Z(\bt_1)Z(\bt_2)\ket_\conn^{g=0}&=e^{-(\bt_1+\bt_2)E_0}\frac{\rt{\bt_1\bt_2}}{2\pi(\bt_1+\bt_2)}\\
&=e^{-(\bt_1+\bt_2)E_0}\int_0^\infty bdb
Z_{\text{trumpet}}(\bt_1,b)Z_{\text{trumpet}}(\bt_2,b)\\
&=\int_0^\infty dE\frac{\psi_E(\bt_1)\psi_{E}(\bt_2)}{2\cosh\pi E}.
\end{aligned} 
\label{eq:psi-wormhole}
\end{equation}
In the last step we used the orthogonality of $\psi_{2E}(2b/\rt{E_0})$.
The last line 
of \eqref{eq:psi-wormhole} agrees with the known result of genus-zero two-point function
\cite{Moore:1991ir} written in terms of the Liouville wavefunction
$\psi_E(\bt)$. 
As discussed in \cite{Moore:1991ir,Moore:1991ag},
the factor $G(E)=(2\cosh\pi E)^{-1}$ in \eqref{eq:psi-wormhole} is interpreted
as the propagator of 2d gravity.
It is interesting to observe that
the square root of the propagator $\rt{2G(E)}=(\cosh\pi E)^{-\hf}$ appears
in the trumpet \eqref{eq:trumpet-wave}.

We can compute various amplitudes
by gluing the trumpet \eqref{eq:trumpet-wave} and $\cM(b)$.
Let us consider the $Z(\bt)$-FZZT amplitude (or half-wormhole)
\begin{equation}
\begin{aligned}
 \bra Z(\bt)|\text{FZZT}\ket&=\int_0^\infty dbZ_{\text{trumpet}}(\bt,b)(-e^{-zb}).
\end{aligned} 
\end{equation}
Using the relation \cite{Martinec:2003ka}
\begin{equation}
\begin{aligned}
 \int_0^\infty\frac{d\bt}{\bt}e^{-\bt E_0\cosh\si}K_{\ri E}(\bt E_0)
=\frac{\pi\cos(E\si)}{E\sinh\pi E},
\end{aligned} 
\label{eq:martinec}
\end{equation}
we find
\begin{equation}
\begin{aligned}
 \bra Z(\bt)|\text{FZZT}(\si)\ket&=-e^{\bt E_0}
\int_0^\infty dE K_{\ri E}(\bt E_0)\frac{\cos(E\si)}{\pi\cosh\pi E},
\end{aligned} 
\label{eq:half-psi}
\end{equation}
where $z$ is related to $\si$ by \eqref{eq:z-si}.
This is written as the error function \eqref{eq:z-HW}.
Appearance of the error function in the $Z(\bt)$-ZZ brane 
amplitude is observed in \cite{Kutasov:2004fg}.

From \eqref{eq:half-psi} we can compute
the annulus amplitude between two FZZT branes
with parameters $z=2\rt{E_0}\cosh\frac{\si}{2}$ and
$z'=2\rt{E_0}\cosh\frac{\si'}{2}$. We find 
\begin{equation}
\begin{aligned}
 \bra\text{FZZT}(\si)|\text{FZZT}(\si')\ket&=-\int_0^\infty\frac{d\bt}{\bt}
e^{-\hf z^2\bt}\bra Z(\bt)|\text{FZZT}(\si')\ket\\
&=\int_0^\infty dE \int_0^\infty\frac{d\bt}{\bt}e^{-\bt E_0\cosh\si}
K_{\ri E}(\bt E_0)\frac{\cos(E\si')}{\pi\cosh\pi E}\\
&=\int_0^\infty dE\frac{2\cos(E\si)\cos(E\si')}{E\sinh2\pi E}.
\end{aligned} 
\end{equation}
The last integral has a divergence coming from $E=0$.
As discussed in \cite{Kutasov:2004fg}, this divergence can be regularized
by taking the principal value
\begin{equation}
\begin{aligned}
 \bra\text{FZZT}(\si)|\text{FZZT}(\si')\ket&=
\int_{-\infty}^\infty dE\frac{E}{E^2+\ve^2}\frac{\cos(E\si)\cos(E\si')}{\sinh2\pi E}\\
&=\frac{1}{2\ve}-\log\left(2\cosh\frac{\si}{2}+2\cosh\frac{\si'}{2}\right)+\cO(\ve).
\end{aligned} 
\end{equation}
After removing the divergent term, the annulus amplitude is written as
\begin{equation}
\begin{aligned}
 \bra\text{FZZT}(\si)|\text{FZZT}(\si')\ket&=-\log(z+z').
\end{aligned} 
\label{eq:annulus}
\end{equation}
This expression agrees with the result of $(2,p)$ minimal string 
\cite{Kutasov:2004fg}.

To summarize, we have shown that the trumpet can be written
in terms of the Liouville wavefunction \eqref{eq:trumpet-wave}
and the known result of annulus amplitude in minimal string theory
is reproduced by gluing
the two trumpets. This justifies our use of trumpet 
in the general background $\{t_k\}$ away from the JT gravity point $t_k=\ga_k$.

We note in passing that the annulus amplitude between two FZZT branes can be
obtained directly by using \eqref{eq:cM-int}
\begin{equation}
\begin{aligned}
 \bra \text{FZZT}(z)|\text{FZZT}(z')\ket&=\int_0^\infty\frac{d\bt}{\bt}e^{-\hf z^2\bt}
\int_0^\infty\frac{d\bt'}{\bt'}e^{-\hf z'^2\bt'}\int_0^\infty bdb Z_{\text{trumpet}}(\bt,b)
Z_{\text{trumpet}}(\bt',b)\\
&=\int_0^\infty \frac{db}{b}e^{-(z+z')b}\\
&=-\log(z+z').
\end{aligned} 
\end{equation}
Here we have ignored the logarithmic divergence coming from $b=0$.
This agrees with \eqref{eq:annulus} as expected.

\section{Deformation of JT gravity}{\label{sec:deform}}

In this section we focus on JT gravity
and study the effect of adding FZZT branes.
We also comment on how it is related to
the deformation of the dilaton potential in JT gravity
that corresponds to adding conical defects.

\subsection{Genus-zero density of states}{\label{sec:rho}}
In this section we consider
$K$ FZZT branes $\det(z^2/2+H)^K$ in JT gravity background and take the
't Hooft limit\footnote{In this section $t$ denotes
the 't Hooft coupling, which
should not be confused with $t=1-I_1$ in \eqref{eq:IZvar}.}
\begin{equation}
\begin{aligned}
 g_s\to0,~~K\to\infty\quad\text{with}~~t=g_sK~~\text{fixed}.
\end{aligned} 
\label{eq:thooft}
\end{equation}
As shown in \eqref{eq:tilde-tk}, adding FZZT branes
amounts to shifting the couplings $t_k$
\begin{equation}
\begin{aligned}
 t_k=\ga_k-t(2k-1)!!z^{-2k-1},
\end{aligned} 
\label{eq:JT-FZZT}
\end{equation}
where $\ga_k$ in \eqref{eq:ga-JT} corresponds to the JT gravity background.
The Itzykson-Zuber variable $I_0(u)$ in the shifted background \eqref{eq:JT-FZZT}
reads
\begin{equation}
\begin{aligned}
f(u):= u-I_0(u)=\sum_{k=0}^\infty (\cob_{k,1}-t_k)\frac{u^k}{k!}
=\rt{u}J_1(2\rt{u})+\frac{t}{\rt{z^2-2u}}.
\end{aligned} 
\label{eq:def-f}
\end{equation}
Here $J_1(x)$ denotes the Bessel function.
As discussed in \cite{Okuyama:2019xbv}, the genus-zero density of states
is determined in terms of the function $f(u)$ in \eqref{eq:def-f}
\begin{equation}
\begin{aligned}
 \rho_0(E)&=\frac{1}{\rt{2}\pi \gs}\int_{E_0}^E\frac{dv}{\rt{E-v}}\frac{\del f(-v)}{\del(-v)}\\
&=\frac{1}{\rt{2}\pi \gs}\int_{E_0}^E\frac{dv}{\rt{E-v}}\left(
I_0(2\rt{v})+\frac{t}{(z^2+2v)^{3/2}}\right)\\
&=\frac{1}{\rt{2}\pi \gs}\left(
\int_{E_0}^E dv\frac{I_0(2\rt{v})}{\rt{E-v}}
+\frac{2t}{z^2+2E}\rt{\frac{E-E_0}{z^2+2E_0}}\right).
\end{aligned} 
\label{eq:rhoE}
\end{equation}
Here $I_0(2\rt{v})$ denotes the modified Bessel function
of the first kind, which should not be confused with the 
Itzykson-Zuber variable $I_0(u)$ in \eqref{eq:def-f}.
The threshold energy $E_0$ is determined by the condition
\begin{equation}
\begin{aligned}
 f(-E_0)=-\rt{E_0}I_1(2\rt{E_0})+\frac{t}{\rt{z^2+2E_0}}=0.
\end{aligned} 
\label{eq:E0-eq}
\end{equation}
When $t=0$, the coupling \eqref{eq:JT-FZZT} reduces to
the JT gravity background $t_k=\ga_k$. Indeed, 
from \eqref{eq:E0-eq} one can see that $E_0=0$ when $t=0$ and 
\eqref{eq:rhoE} reproduces the known result of JT gravity density
of states \cite{Stanford:2017thb}
\begin{equation}
\begin{aligned}
 \rho_0(E)=\frac{1}{\rt{2}\pi \gs}\int_{0}^E dv\frac{I_0(2\rt{v})}{\rt{E-v}}
=\frac{\sinh(2\rt{E})}{\rt{2}\pi \gs}.
\end{aligned} 
\label{eq:JT-rho}
\end{equation}

\begin{figure}[htb]
\centering
\includegraphics[width=0.7\linewidth]{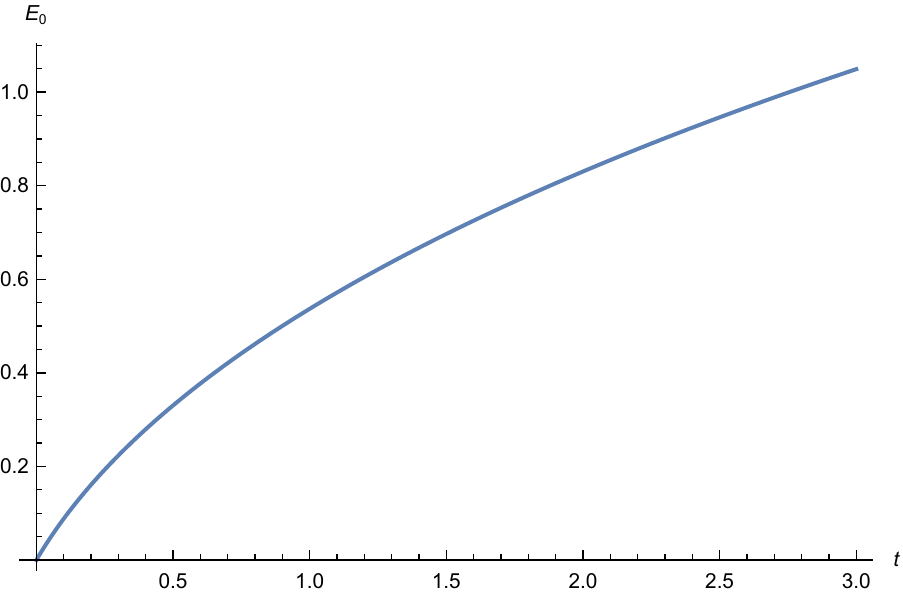}
  \caption{
Plot 
of the threshold energy $E_0$ as a function of $t=\gs K$. 
We set $z=1$ in this figure.
}
  \label{fig:E0}
\end{figure}

\begin{figure}[htb]
\centering
\includegraphics[width=0.7\linewidth]{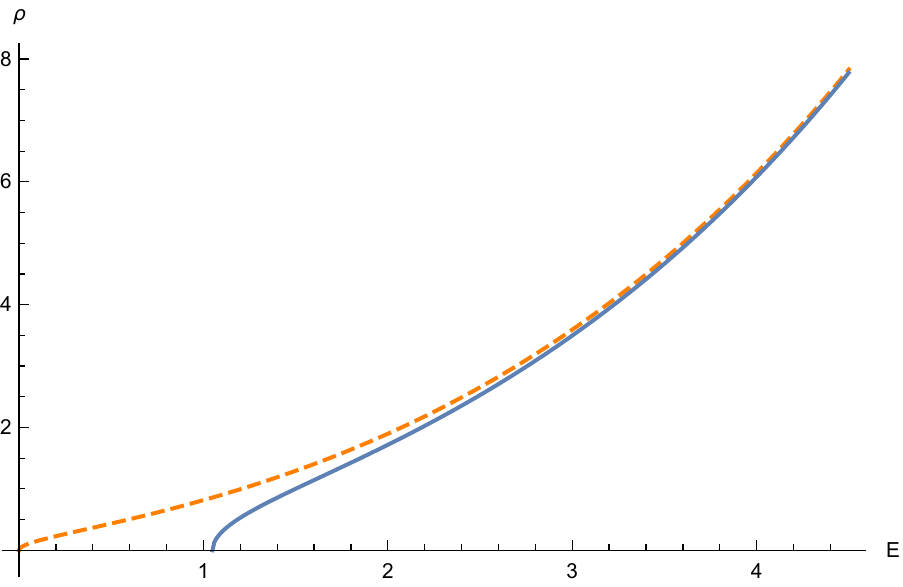}
  \caption{
Plot 
of the eigenvalue density $\rho_0(E)$
for $t=3,z=1,g_s=1$ (blue solid curve). The orange dashed curve
represents the pure JT gravity density of states
in \eqref{eq:JT-rho}.
}
  \label{fig:rho}
\end{figure}

For $t\ne0$, we do not have a closed form expression
of the solution $E_0$
of \eqref{eq:E0-eq}.
However, one can easily find the small $t$ and large $t$ behavior
of $E_0$.
From \eqref{eq:E0-eq}, one can show that
in the small $t$ regime $E_0$ is expanded as
\begin{equation}
\begin{aligned}
 E_0=\frac{t}{z}-\left(\frac{1}{z^4}+\frac{1}{2z^2}\right)t^2
+\left(\frac{5}{2z^7}+\frac{3}{2z^5}+\frac{5}{12z^3}\right)t^3+\cO(t^4).
\end{aligned} 
\end{equation}
On the other hand,
in the large $t$ regime $E_0$ is large and \eqref{eq:E0-eq} is approximated as
\begin{equation}
\begin{aligned}
 -\rt{E_0}\frac{e^{2\rt{E_0}}}{\rt{4\pi\rt{E_0}}}+\frac{t}{\rt{2E_0}}=0.
\end{aligned} 
\end{equation}
This is rewritten as
\begin{equation}
\begin{aligned}
 \frac{4}{3}(2\pi t^2)^{1/3}=\frac{4}{3}\rt{E_0}\,e^{\frac{4}{3}\rt{E_0}}.
\end{aligned} 
\end{equation}
The solution of this equation
is given by the Lambert $W$-function obeying $z=W(z)e^{W(z)}$
\begin{equation}
\begin{aligned}
 E_0=\left[\frac{3}{4}W\left(\frac{4}{3}(2\pi t^2)^{1/3}\right)\right]^2.
\end{aligned} 
\end{equation}
In the intermediate value of $t$, one can solve the equation
\eqref{eq:E0-eq} numerically.
In Figure~\ref{fig:E0}, we show the plot of $E_0$ as a function of $t$.
Once we know $E_0$, we can numerically evaluate the integral
\eqref{eq:rhoE} to find $\rho_0(E)$. 
In Figure~\ref{fig:rho}, we show the plot of $\rho_0(E)$ for 
$t=3,z=1,\gs=1$ as an example. 
As a comparison, we also plot the pure JT gravity density of states
\eqref{eq:JT-rho}
(see orange dashed curve in Figure~\ref{fig:rho}).

When $z^2>0$, the eigenvalue corresponding to the FZZT brane is located at $E=-z^2/2<0$.
Since the eigenvalues behave as fermions and they repel each other,
the other eigenvalues are pushed toward the positive $E$ direction,
as we can see from Figure~\ref{fig:rho}.

On the other hand, if $E=-z^2/2>0$ the corresponding eigenvalue
is inserted on the original support of the pure JT gravity eigenvalue density \eqref{eq:JT-rho}.
We expect this insertion creates a void in the support of $\rho_0(E)$.
As we increase $t$, we expect that the support of eigenvalue density is 
split into two parts and the model exhibits a phase transition from a one-cut phase 
to a two-cut phase.
However, in our formalism \eqref{eq:FZZT-formula} of the perturbative
treatment of FZZT branes we assumed $\text{Re}(z^2)>0$
for the derivation of $\cM(b)=-e^{-zb}$ in \eqref{eq:cMb}.
In other words, our formalism \eqref{eq:FZZT-formula} is valid only for the
one-cut phase of the matrix model.
To see the effect of FZZT branes with $\text{Re}(z^2)<0$
we need a non-perturbative treatment of FZZT branes, which we will consider
in section~\ref{sec:corr}.

Before closing this subsection,
let us compute the order $\cO(t^1)$ correction of the density of states
\eqref{eq:rhoE}. One can show that
\begin{equation}
\begin{aligned}
 \rho_0(E)\Big|_{\cO(t)}&=\frac{1}{\rt{2}\pi\gs}\left(-\frac{t}{z\rt{E}}+
\frac{2t\rt{E}}{(z^2+2E)z}\right)\\
&=-\frac{t}{\rt{2}\pi\gs}\frac{z}{(z^2+2E)\rt{E}}.
\end{aligned} 
\end{equation}
Then the order $\cO(t)$ correction to the one-point function $\bra Z(\bt)\ket$
is given by
\begin{equation}
\begin{aligned}
 \int_0^\infty dE e^{-\bt E}\rho_0(E)\Big|_{\cO(t)}&=-\frac{t}{2\gs}
e^{\hf z^2\bt}\text{Erfc}\left(\rt{\frac{z^2\bt}{2}}\right)
=KZ_{\text{HW}}(\bt,z),
\end{aligned} 
\label{eq:rho-t}
\end{equation}
where $Z_{\text{HW}}(\bt,z)$
is the ``half-wormhole'' amplitude in \eqref{eq:z-HW}.
In \eqref{eq:rho-t} $Z_{\text{HW}}(\bt,z)$ is
multiplied by a factor of $K$
since we are considering $K$ FZZT branes.
Thus our expression of the genus-zero density of states
in \eqref{eq:rhoE} correctly reproduces the half-wormhole contribution 
in \eqref{eq:z-HW}.

\subsection{String equation via the Lagrange reversion}
As discussed in \cite{Maxfield:2020ale}, the modification of
the genus-zero string equation $f(u)=0$
with $f(u)$ in \eqref{eq:def-f} can also be seen by
using the Lagrange reversion theorem.
Following \cite{Maxfield:2020ale}, let us consider the
modification of the genus-zero part of the two-point function
$\bra Z(\bt_1)Z(\bt_2)\ket_\conn^{g=0}$
due to the insertion of $K$ FZZT branes $\det(z^2/2+H)^K$.
By using our formalism \eqref{eq:FZZT-formula}
in the general background $\{t_k\}$, the genus-zero
two-point function
in the presence of $K$ FZZT branes is written as
\begin{equation}
\begin{aligned}
 &\bra Z(\bt_1)Z(\bt_2)\det(z^2/2+H)^K\ket_\conn^{g=0}\\
=&\frac{\rt{\bt_1\bt_2}}{2\pi(\bt_1+\bt_2)}e^{(\bt_1+\bt_2)u_0}\\
&+\sum_{n=1}^\infty\frac{\gs^{n}}{n!}
\prod_{i=1}^2\int_0^\infty b_i'db_i'Z_{\text{trumpet}}(\bt_i,b_i')
\prod_{j=1}^n\int_0^\infty  db_j\cM(b_j)
V_{0,n+2}(b_1',b_2',b_1,\cdots,b_n),
\end{aligned} 
\label{eq:two-pt-deform}
\end{equation}
where $V_{0,n+2}(\bm{b})$ is given by \eqref{eq:V0n} and $\cM(b)$ for $K$ FZZT branes
is given by
\begin{equation}
\begin{aligned}
 \cM(b)=-Ke^{-zb}.
\end{aligned} 
\end{equation} 
By using \eqref{eq:trumpet-Iint} and \eqref{eq:cM-Ik-int},
\eqref{eq:two-pt-deform} is written as
\begin{equation}
\begin{aligned}
 &\bra Z(\bt_1)Z(\bt_2)\det(z^2/2+H)^K\ket_\conn^{g=0}\\
=&\frac{\rt{\bt_1\bt_2}}{2\pi(\bt_1+\bt_2)}e^{(\bt_1+\bt_2)u_0(t_0)}\\
&+\frac{\rt{\bt_1\bt_2}}{2\pi}
\sum_{n=1}^\infty\frac{(-t)^{n}}{n!}\del_0^{n-1}\Bigl(
\del_0u_0(t_0) e^{(\bt_1+\bt_2)u_0(t_0)}(z^2-2u_0(t_0))^{-\frac{n}{2}}\Bigr)\\
=&\frac{\rt{\bt_1\bt_2}}{2\pi(\bt_1+\bt_2)}\left[
e^{(\bt_1+\bt_2)u_0(t_0)}+\sum_{n=1}^\infty\frac{(-t)^n}{n!}
\del_0^{n-1}\Bigl(\del_0 e^{(\bt_1+\bt_2)u_0(t_0)}
(z^2-2u_0(t_0))^{-\frac{n}{2}}\Bigr)
\right],
\end{aligned} 
\label{eq:2pt-reversion}
\end{equation}
where $t_0$-dependence has been explicitly 
expressed for convenience of explanation.
Let us now recall the Lagrange reversion theorem
(see e.g.~\cite{Whittaker}):
Suppose that $\zeta$ is related to $a$ by the equation
\begin{align}
\zeta=a+t\phi(\zeta)
\end{align}
with $\phi(z)$ being some function.
Then for any function $g(z)$ and for small enough $t$,
$g(\zeta)$ is expanded as 
\begin{align}
g(\zeta)=g(a)+\sum_{n=1}^\infty\frac{t^n}{n!}\frac{d^{n-1}}{da^{n-1}}
\left[g'(a)\{\phi(a)\}^n\right].
\end{align}
Using this theorem one immediately sees that
\eqref{eq:2pt-reversion} is rewritten as
\begin{align}
\bra Z(\bt_1)Z(\bt_2)\det(z^2/2+H)^K\ket_\conn^{g=0}
=\frac{\rt{\bt_1\bt_2}}{2\pi(\bt_1+\bt_2)}e^{(\bt_1+\bt_2)u_0(x)}
\label{eq:2pt-reversion2}
\end{align}
with
\begin{align}
x=t_0-\frac{t}{\sqrt{z^2-u_0(x)}}.
\label{eq:xtrel}
\end{align}
Now,
the genus zero string equation $u_0-I_0(u_0)=0$ is written
in the off-shell JT gravity background $t_k=\gamma_k\ (k\ge 1)$ as
\begin{align}
0=\rt{u_0(t_0)}J_1\left(2\rt{u_0(t_0)}\right)-t_0.
\end{align}
Writing this equation at $t_0=x$ with $x$ in \eqref{eq:xtrel}
and then setting $t_0=0$ to
restrict ourselves to the on-shell JT gravity background,
we have
\begin{align}
\begin{aligned}
0&=\rt{u_0(x)}J_1\left(2\rt{u_0(x)}\right)-x\\
 &=\rt{u_0(x)}J_1\left(2\rt{u_0(x)}\right)+\frac{t}{\sqrt{z^2-u_0(x)}}.
\end{aligned}
\label{eq:modstreq}
\end{align}
We have thus seen that the genus zero two-point function 
in the presence of $K$ FZZT-branes is written as
the genuine genus zero two-point function \eqref{eq:2pt-reversion2}
with $u_0(x)$ satisfying the equation \eqref{eq:modstreq}.
This is identical with the modified string equation $f(u_0(x))=0$
with $f(u)$ given in \eqref{eq:def-f},
which was obtained from the shift of couplings \eqref{eq:JT-FZZT}.

\subsection{Comment on dilaton gravity}

In \cite{Witten:2020wvy,Maxfield:2020ale}, it is shown that
we can add conical defects in JT gravity by
modifying the dilaton potential
\begin{equation}
\begin{aligned}
 S=-\hf \int d^2x\rt{g}\Bigl[\phi(R+2)+U(\phi)\Bigr].
\end{aligned} 
\end{equation}
For the defects with deficit angle $2\pi\al_i$ with fugacity $\ve_i$,  
$U(\phi)$ is given by
\begin{equation}
\begin{aligned}
 U(\phi)=2\sum_i\ve_i e^{-2\pi(1-\al_i)\phi}.
\end{aligned} 
\label{eq:Uphi}
\end{equation}
As discussed in \cite{Maxfield:2020ale} the effect of defects
is summarized as the modification of the string equation (see also 
\cite{Budd,Johnson:2020lns,Forste:2021roo})\footnote{This is obtained
as follows. The correction to the one-point function of
$Z(\bt)$ due to defects is given by \cite{Mertens:2019tcm}
\begin{equation}
\begin{aligned}
 \sum_i\ve_i Z(\bt,\al_i)=\sum_i\ve_i\frac{e^{\frac{\al^2_i}{\bt}}}{\rt{2\pi\bt}}.
\end{aligned} 
\end{equation}
This is formally equal to the trumpet partition function
with the replacement $b=\rt{2}\ri\al_i$. Using the Lagrange reversion theorem 
one can show that the Itzykson-Zuber variable $I_0(u_0)$ is shifted
by
\begin{equation}
\begin{aligned}
 \sum_i\ve_i\cI_0(\rt{2}\ri\al_i)=\sum_i\ve_iJ_0(2\al_i\rt{u_0}),
\end{aligned} 
\end{equation}
where $\cI_0(b)$ is defined in \eqref{eq:cI-def}. Thus we find \eqref{eq:string-defect}. 
}
\begin{equation}
\begin{aligned}
 u-I_0(u)=\rt{u}J_1(2\rt{u})-\sum_i\ve_i J_0(2\al_i\rt{u}).
\end{aligned}
\label{eq:string-defect} 
\end{equation}
From this we can read off the shifted coupling as
\begin{equation}
\begin{aligned}
 t_k=\ga_k+\sum_i\ve_i\frac{(-1)^k\al_i^{2k}}{k!}.
\end{aligned} 
\label{eq:tshift-al}
\end{equation}
Let us consider the relation between this 
shifted coupling and the dilaton potential $U(\phi)$.
To see this, it is convenient to define
\begin{equation}
\begin{aligned}
 \til{U}(\phi)&=U(\phi)e^{2\pi\phi},\quad
\til{U}_+(\phi)&=\frac{\til{U}(\phi)+\til{U}(-\phi)}{2}.
\end{aligned} 
\end{equation}
For the dilaton potential in \eqref{eq:Uphi} we find
\begin{equation}
\begin{aligned}
 \til{U}_+(\phi)=2\sum_i\ve_i\cosh(2\pi\al_i\phi).
\end{aligned} 
\end{equation}
From \eqref{eq:tshift-al}, this is also written as
\begin{equation}
\begin{aligned}
 \til{U}_+(\phi)&=2\sum_{k=0}^\infty
\frac{(-1)^k(t_k-\ga_k)}{(2k-1)!!}\frac{(2\pi\phi)^{2k}}{2^k}.
\end{aligned} 
\label{eq:tilU-gen}
\end{equation}
As argued in \cite{Witten:2020wvy}, the equivalence of
the matrix model and dilaton gravity holds 
for a wide class of dilaton potential written as 
a superposition of the defect potential \eqref{eq:Uphi}
\begin{equation}
\begin{aligned}
 U(\phi)=2\int d\al \ve(\al)e^{-2\pi(1-\al)\phi}.
\end{aligned} 
\end{equation}
Thus we expect that \eqref{eq:tilU-gen} can be 
applied to a wide class of background couplings $\{t_k\}$.
For instance, if we add $K$ FZZT branes the coupling is shifted as \eqref{eq:JT-FZZT}.
Plugging \eqref{eq:JT-FZZT} into \eqref{eq:tilU-gen},
the corresponding dilaton potential becomes
\begin{equation}
\begin{aligned}
 \til{U}_+(\phi)=-\frac{2tz}{z^2+2\pi^2\phi^2}.
\end{aligned} 
\end{equation}
In this approach we can only find the even part of $\til{U}(\phi)=U(\phi)e^{2\pi\phi}$.
We do not know how to recover the odd part of $\til{U}(\phi)$
from the data of $\{t_k\}$.

Also, it is not clear to us what is the condition for the applicability of the 
formula \eqref{eq:tilU-gen}.
We do not know how far we can deform the coupling from the JT gravity background
$t_k=\ga_k$ when we use \eqref{eq:tilU-gen}. Putting this
problem aside, if we naively apply \eqref{eq:tilU-gen}
to the $(2,p)$ minimal model background
\eqref{eq:tk-min} we find
\begin{equation}
\begin{aligned}
 \til{U}_+(\phi)=2\pi\phi\sinh\left[p\,
\text{arcsinh}\left(\frac{2\pi\phi}{p}\right)\right]-2\pi\phi\sinh(2\pi\phi).
\end{aligned} 
\label{eq:tilU-min}
\end{equation}
In the $p\to\infty$ limit, $\til{U}_+(\phi)$ in \eqref{eq:tilU-min} vanishes, which
is consistent with the statement that JT gravity is a $p\to\infty$ limit
of the $(2,p)$ minimal string \cite{Seiberg:2019, Saad:2019lba}.

In \cite{Seiberg:2019,Mertens:2020hbs,Turiaci:2020fjj}, it is proposed that
a dilaton gravity theory with $\sinh\phi$
potential corresponds to the $(2,p)$ minimal string.
It would be interesting to understand the relation
between our \eqref{eq:tilU-min} and the proposal in \cite{Seiberg:2019,Mertens:2020hbs,Turiaci:2020fjj}, if any.

\section{Baker-Akhiezer function and FZZT branes}\label{sec:BA}
In this section we review the known results
about the Baker-Akhiezer (BA) function,
the Christoffel-Darboux (CD) kernel and
the multi-FZZT amplitude.

\subsection{Multi-FZZT amplitude from matrix integral}
As discussed in \cite{Maldacena:2004sn},
the BA function can be thought of as the wavefunction
of FZZT brane, which is obtained as a double-scaling limit of the
orthogonal polynomial.
In the matrix model at finite $N$
\begin{equation}
\begin{aligned}
 Z_N=\int_{N\times N} dM e^{-\Tr V(M)},
\end{aligned} 
\label{eq:ZN}
\end{equation}
it is convenient to define the polynomial $P_n(\la)=\la^n+\cdots$ 
of degree $n$ which is orthogonal with respect to the weight $e^{-V(\la)}$
\begin{equation}
\begin{aligned}
 \int d\la e^{-V(\la)}P_n(\la)P_m(\la)=h_n\cob_{n,m}.
\end{aligned} 
\label{eq:Pn-orth}
\end{equation}
One can show that $Z_N$ in \eqref{eq:ZN} is written in terms of the norm $h_n$
of the orthogonal polynomial $P_n(\la)$
\begin{equation}
\begin{aligned}
 Z_N=\prod_{n=0}^{N-1}h_n.
\end{aligned}
\label{eq:ZN-norm} 
\end{equation}
It is convenient to define the normalized orthogonal polynomial by
\begin{equation}
\begin{aligned}
 \psi_n(\la)=\frac{e^{-\hf V(\la)}}{\rt{h_n}}P_n(\la),
\end{aligned} 
\end{equation}
which satisfies
\begin{equation}
\begin{aligned}
 \int d\la\psi_n(\la)\psi_m(\la)=\cob_{n,m}.
\end{aligned} 
\end{equation}
It is well known that $\psi_N(\la)$ is given by the expectation value of the determinant
operator in the matrix model (see e.g.~\cite{Eynard:2015aea} for a review)
\begin{equation}
\begin{aligned}
 \psi_N(\la)&=\bra\Psi_N(\la)\ket=\frac{1}{Z_N}\int_{N\times N}dM e^{-\Tr V(M)}\Psi_N(\la),
\end{aligned} 
\end{equation}
where $\Psi_N(\la)$ is given by
\begin{equation}
\begin{aligned}
 \Psi_N(\la)=\frac{e^{-\hf V(\la)}}{\rt{h_N}}\det_{N\times N}(\la-M).
\end{aligned} 
\end{equation}
Another important ingredient is the CD kernel 
\begin{equation}
\begin{aligned}
 K_N(\xi,\eta)&=\sum_{n=0}^{N-1}\psi_n(\xi)\psi_n(\eta).
\end{aligned} 
\label{eq:CD-N}
\end{equation}
Using the relation
\begin{equation}
\begin{aligned}
 \la\psi_n(\la)=\rt{r_{n+1}}\psi_{n+1}(\la)+\rt{r_n}\psi_{n-1}(\la),\quad
r_n=\frac{h_n}{h_{n-1}},
\end{aligned}
\label{eq:jacobi-mat} 
\end{equation}
the CD kernel is written as
\begin{equation}
\begin{aligned}
 K_N(\xi,\eta)=
\rt{r_N}\frac{\psi_N(\xi)\psi_{N-1}(\eta)-\psi_{N-1}(\xi)\psi_{N}(\eta)}{\xi-\eta}.
\end{aligned} 
\label{eq:CDN-frac}
\end{equation}
One can show that
the CD kernel \eqref{eq:CD-N} is given by the two-point function of
determinant operators
\begin{equation}
\begin{aligned}
 K_{N}(\xi,\eta)=\bra\Psi_{N-1}(\xi)\Psi_{N-1}(\eta)\ket,
\end{aligned} 
\end{equation}
where the expectation value is taken in the $(N-1)\times (N-1)$ matrix integral. 
In the double-scaling limit the index $n$ of $\psi_n(\la)$
becomes a continuous coordinate $t_0$
\begin{equation}
\begin{aligned}
\psi_n(\la)\quad\to\quad\psi(\lambda;t_0),
\end{aligned} 
\end{equation}
and the double-scaling limit of \eqref{eq:CD-N} becomes
\begin{equation}
\begin{aligned}
K(\xi,\eta)
 =\frac{1}{\hbar}\int_{-\infty}^{t_0}dx\psi(\xi;x)\psi(\eta;x).
\end{aligned} 
\label{eq:CDkernel}
\end{equation}
Here $\hbar$ is related to the genus counting parameter $\gs$ by
\begin{equation}
\begin{aligned}
 \hbar=\frac{\gs}{\rt{2}}.
\end{aligned} 
\end{equation}
From \eqref{eq:CDkernel}, one can see that $K(\xi,\eta)$ satisfies
\begin{equation}
\begin{aligned}
 \hbar\del_0K(\xi,\eta)=\psi(\xi)\psi(\eta),
\end{aligned} 
\label{eq:del-K}
\end{equation}
where $\psi(\xi)=\psi(\xi;t_0)$.
One can also show that \eqref{eq:jacobi-mat} reduces to the Schr\"{o}dinger
equation in the double-scaling limit (see \cite{Ginsparg:1993is} for a review)
\begin{equation}
\begin{aligned}
 \big(\hbar^2\del_0^2+u\big)\psi(\la)=\la\psi(\la),
\end{aligned}
\label{eq:Sch} 
\end{equation}
where $u$ is the specific heat
\begin{equation}
\begin{aligned}
 u=\gs^2\del_0^2F=2\hbar^2\del_0^2F.
\end{aligned} 
\label{eq:udef}
\end{equation}
Note that 
the CD kernel of the form \eqref{eq:CDN-frac} in the double scaling limit becomes
\begin{equation}
\begin{aligned}
 K(\xi,\eta)=\hbar\frac{\del_0\psi(\xi)\psi(\eta)-\psi(\xi)\del_0\psi(\eta)}{\xi-\eta}.
\end{aligned} 
\end{equation}
This is also derived from \eqref{eq:CDkernel} by using \eqref{eq:Sch}.

As discussed in \cite{Maldacena:2004sn} the multi-point function of
FZZT branes can be obtained by taking the double scaling limit of 
the finite $N$ expression \cite{Morozov:1994hh,Brezin:2000}
\begin{equation}
\begin{aligned}
 \left\bra\prod_{i=1}^k\det(\xi-M)\right\ket_{N\times N}=\frac{1}{\lap(\xi)}
\det\Bigl(P_{N+i-1}(\xi_j)\Bigr)_{i,j=1,\ldots,k},
\end{aligned} 
\end{equation}
where $\lap(\xi)=\prod_{i<j}(\xi_i-\xi_j)$ is the Vandermonde determinant.
In the double scaling limit one finds 
\cite{Maldacena:2004sn}
\begin{equation}
\begin{aligned}
 \left\bra\prod_{i=1}^k\Psi(\xi)\right\ket=\frac{\lap(d)}{\lap(\xi)}\prod_{i=1}^k
\psi(\xi_i).
\end{aligned} 
\label{eq:lap-d}
\end{equation}
Here $d_j$ is the shorthand for the action of $d=\hbar\del_0$ on 
$\psi(\xi_j)$.

\subsection{Genus expansion of BA function} 
Let us consider the genus expansion of the BA function
$\psi(\xi)=\psi(\xi;t_0)$.
It is known that the BA function is written as a ratio
of tau-function $\tau(\bm{t})=e^{F(\bm{t})}$
\cite{Date:1982yeu,dubrovin}:
\begin{equation}
\begin{aligned}
 \psi(\xi)=e^{\vartheta(z)}\frac{\tau(\bm{t}-[z^{-1}])}{\tau(\bm{t})}
=e^{\vartheta(z)+F(\bm{t}-[z^{-1}])-F(\bm{t})},
\end{aligned} 
\label{eq:BA-tau}
\end{equation}
where $\xi=z^2/2$ and
\begin{equation}
\begin{aligned}
 \vartheta(z)=\frac{1}{\gs}\sum_{k=0}^\infty \frac{t_k-\cob_{k,1}}{(2k+1)!!}z^{2k+1}.
\end{aligned} 
\end{equation} 
$[z^{-1}]$ in \eqref{eq:BA-tau} is defined by
\begin{equation}
\begin{aligned}
{}[z^{-1}]_k=\gs (2k-1)!!z^{-2k-1}.
\end{aligned} 
\end{equation}
This is consistent with our result of the shift of couplings \eqref{eq:tilde-tk}.

Let us consider the genus expansion of the BA function
\begin{equation}
\begin{aligned}
 \log\psi(\xi)=\vartheta(z)+F(\bm{t}-[z^{-1}])-F(\bm{t})=\sum_{n=0}^\infty\gs^{n-1}A_n.
\end{aligned} 
\label{eq:logpsi}
\end{equation}
The $A_0$ term is known as the effective potential
\begin{equation}
\begin{aligned}
 A_0=-\hf V_{\text{eff}}(\xi),
\end{aligned} 
\end{equation}
whose explicit form is obtained in \cite{Okuyama:2020ncd} as
\begin{equation}
\begin{aligned}
  \hf V_{\text{eff}}(\xi)=\sum_{n=1}^\infty 
\frac{\cob_{n,1}-I_n}{(2n+1)!!}(z^2-2u_0)^{n+\hf}.
\end{aligned} 
\label{eq:Veff}
\end{equation}
On the other hand, from \eqref{eq:logpsi} we find
\begin{equation}
\begin{aligned}
  \hf V_{\text{eff}}(\xi)=\sum_{k=0}^\infty
\left[\frac{\cob_{k,1}-t_k}{(2k+1)!!}z^{2k+1}+\frac{(2k-1)!!}{z^{2k+1}}\del_kF_0
\right].
\end{aligned}
\label{eq:Veff-z} 
\end{equation}
We can show the equivalence of \eqref{eq:Veff} and \eqref{eq:Veff-z} as follows.
It turns out that  \eqref{eq:Veff} has the following integral representation
\begin{equation}
\begin{aligned}
  \hf V_{\text{eff}}(\xi)=\int_{u_0}^{z^2/2}du\frac{u-I_0(u)}{\rt{z^2-2u}}.
\end{aligned} 
\label{eq:Veff-int}
\end{equation}
This can be shown by using the relations
\begin{equation}
\begin{aligned}
 \del_u I_k(u)=I_{k+1}(u),\quad
u_0-I_0(u_0)=0
\end{aligned} 
\end{equation}
and repeating the partial integration:
\begin{equation}
\begin{aligned}
 \hf V_{\text{eff}}(\xi)&= 
-(u-I_0(u))(z^2-2u)^{\frac{1}{2}}\Big|_{u=u_0}^{u=\frac{z^2}{2}}
+\int_{u_0}^{z^2/2}du (1-I_1(u))(z^2-2u)^{\frac{1}{2}} \\
&=-\frac{1-I_1(u)}{3}(z^2-2u)^{\frac{3}{2}}\Big|_{u=u_0}^{u=\frac{z^2}{2}}
-\int_{u_0}^{z^2/2}du\frac{I_2(u)}{3}(z^2-2u)^{\frac{3}{2}}\\
&=\frac{1-I_1(u_0)}{3}(z^2-2u_0)^{\frac{3}{2}}
+\frac{I_2(u)}{5!!}(z^2-2u)^{\frac{5}{2}}\Big|_{u=u_0}^{u=\frac{z^2}{2}}
-\int_{u_0}^{z^2/2}du\frac{I_3(u)}{5!!}(z^2-2u)^{\frac{5}{2}}\\
&=\cdots=\sum_{n=1}^\infty \frac{\cob_{n,1}-I_n}{(2n+1)!!}(z^2-2u_0)^{n+\hf}.
\end{aligned} 
\end{equation} 
Now, let us decompose \eqref{eq:Veff-int} into two parts
\begin{equation}
\begin{aligned}
 \hf V_{\text{eff}}(\xi)&=
\int_{0}^{z^2/2}du\frac{u-I_0(u)}{\rt{z^2-2u}}-
\int_{0}^{u_0}du\frac{u-I_0(u)}{\rt{z^2-2u}}
\end{aligned} 
\label{eq:Veff-decomp}
\end{equation}
One can show that the first term of \eqref{eq:Veff-decomp}
agrees with the first term of \eqref{eq:Veff-z}
\begin{equation}
\begin{aligned}
 \int_{0}^{z^2/2}du\frac{u-I_0(u)}{\rt{z^2-2u}}&=
\int_{0}^{z^2/2}du \sum_{k=0}^\infty
(\cob_{k,1}-t_k)\frac{u^{k}}{k!}(z^2-2u)^{-\hf}\\
&=\sum_{k=0}^\infty\frac{\cob_{1,k}-t_k}{(2k+1)!!}z^{2k+1}.
\end{aligned} 
\end{equation}
Using the relation \cite{Itzykson:1992ya}
\begin{equation}
\begin{aligned}
 F_0=\hf\int_0^{u_0}du(I_0(u)-u)^2,
\end{aligned} 
\end{equation}
one can show that the second term of \eqref{eq:Veff-decomp}
is equal to the second term of \eqref{eq:Veff-z}. Thus we  find
the equivalence of 
\eqref{eq:Veff} and \eqref{eq:Veff-z}.

We can compute the higher genus corrections
of the genus expansion \eqref{eq:logpsi} by introducing the operator 
\begin{equation}
\begin{aligned}
 D(z)=-\sum_{k=0}^\infty (2k-1)!!z^{-2k-1}\del_k.
\end{aligned} 
\end{equation}
Then we find
\begin{equation}
\begin{aligned}
 \log\psi=\vartheta(z) +\sum_{g=0}^\infty\sum_{n=1}^\infty g_s^{2g-2+n}
\frac{D(z)^n}{n!}F_g. 
\end{aligned} 
\end{equation}
For instance, the first few terms are given by
\begin{equation}
\begin{aligned}
 A_0&=\gs\vartheta(z)+D(z)F_0,\\
A_1&=\hf D(z)^2F_0,\\
A_2&=\frac{1}{6}D(z)^3F_0+D(z)F_1.
\end{aligned} 
\end{equation}
In this computation, the following relations are useful
\begin{equation}
\begin{aligned}
 D(z)I_k(u)&=-\frac{(2k-1)!!}{(z^2-2u)^{k+\hf}},\quad
D(z)u_0=-\frac{1}{t\til{z}},
\end{aligned} 
\label{eq:DI-Du0}
\end{equation}
where
\begin{equation}
\begin{aligned}
 \til{z}=\rt{z^2-2u_0}.
\end{aligned} 
\end{equation}
From \eqref{eq:DI-Du0} and $F_1=-\frac{1}{24}\log t$ one can show that
\begin{equation}
\begin{aligned}
 A_1&=-\hf\log\til{z}+\hf\log z,\\
A_2
&=-\frac{5}{24t\til{z}^3}-\frac{I_2}{24t^2\til{z}}.
\end{aligned} 
\end{equation}
This reproduces the known result of the genus expansion
of BA function \cite{Okuyama:2020ncd}
(up to the normalization), as expected.

\subsection{Genus expansion of the multi-FZZT amplitude}
The multi-FZZT amplitude is written as
\begin{equation}
\begin{aligned}
 \left\bra\prod_i\Psi(\xi_i)\right\ket=\frac{\lap(z)}{\lap(\xi)}
e^{\sum_i\vartheta(z_i)+F(\bm{t}-\sum_i[z_i^{-1}])-F(\bm{t})}\equiv
\frac{\lap(z)}{\lap(\xi)}e^{A(z_i)}
\end{aligned} 
\label{eq:multi-A}
\end{equation}
The genus expansion of $A(z_i)$ can be obtained as follows.
The Itzykson-Zuber variable for $\til{t}_k$ in
\eqref{eq:tilde-tk} is written as
\begin{equation}
\begin{aligned}
 \til{I}_k(\til{u}_0)=\sum_{n=0}^\infty \til{t}_{k+n}\frac{\til{u}_0^n}{n!}
=I_k(\til{u}_0)-\gs(2k-1)!!\sum_i (z^2_i-2\til{u}_0)^{-k-\hf}.
\end{aligned} 
\end{equation}
The string equation for the shifted background $\til{t}_k$ becomes
\begin{equation}
\begin{aligned}
 \til{u}_0-I_0(\til{u}_0)+\sum_i\frac{\gs}{\rt{z^2_i-2\til{u}_0}}=0.
\end{aligned} 
\label{eq:til-string}
\end{equation}
From \eqref{eq:til-string} we find the genus expansion of $\til{u}_0$
\begin{equation}
\begin{aligned}
 \til{u}_0=u_0-\sum_i \frac{\gs}{t\til{z}_i}+\cdots,\qquad
\til{z}_i=\rt{z^2_i-2u_0}.
\end{aligned} 
\label{eq:til-u0}
\end{equation}
Then the genus expansion of $F(\bm{t}-\sum_i[z_i^{-1}])$ is obtained by replacing $I_k(u_0)$
in $F(\bm{t})$ by $\til{I}_k(\til{u}_0)$.\footnote{In \cite{Okuyama:2020qpm}
it is shown that the expression of the genus expansion is universal to
the tau-function of the KdV hierarchy when written in terms of
the generalized Itzykson-Zuber variables $\{\til{I}_n\}$.
Replacing $I_k(u_0)$ in $F(\bm{t})$ by $\til{I}_k(\til{u}_0)$
here is equivalent to setting
$\varphi(u_0)=u_0+\displaystyle\sum_i\dfrac{\gs}{\rt{z^2_i-2u_0}}$
in the formalism of \cite{Okuyama:2020qpm}.}
Thus we have
\begin{equation}
\begin{aligned}
 A(z_i)=\sum_i\vartheta(z_i)+F(\til{I}_k(\til{u}_0))-F(I_k(u_0)).
\end{aligned} 
\end{equation}
Recall that $F_g(t)~(g\geq2)$ are
given by polynomials
of $I_k(u_0)~(k\geq2)$
and $(1-I_1(u_0))^{-1}$ \cite{Itzykson:1992ya}.
For $g=0$ and $g=1$ we have
\begin{equation}
\begin{aligned}
 F_0(\bm{t}-\sum_i[z_i^{-1}])&=\hf\int_0^{\til{u}_0} du\left[I_0(u)-\sum_i\frac{\gs}{\rt{z^2_i-2u}}-u\right]^2,\\
F_1(\bm{t}-\sum_i[z^{-1}])&=-\frac{1}{24}\log\left(1-I_0(\til{u}_0)
+\sum_i\frac{\gs}{(z^2_i-2\til{u}_0)^{3/2}}\right).
\end{aligned} 
\end{equation}
Expanding these expressions using \eqref{eq:til-u0}, we can compute the genus
expansion of $A(z_i)$ in \eqref{eq:multi-A}.

\subsection{Annulus amplitude between two FZZT branes}
Let us consider the annulus amplitude between two FZZT branes.
From \eqref{eq:multi-A} we have
\begin{equation}
\begin{aligned}
 \bra\Psi(\xi_1)\Psi(\xi_2)\ket=\frac{z_1-z_2}{\xi_1-\xi_2}e^{A(z_1,z_2)}=\frac{2}{z_1+z_2}e^{A(z_1,z_2)}.
\end{aligned} 
\label{eq:z-ann} 
\end{equation}
The prefactor agrees with the exponentiated
annulus amplitude \eqref{eq:annulus} up to an overall
normalization constant. On the other hand, from \eqref{eq:lap-d} we find
\begin{equation}
\begin{aligned}
 \bra\Psi(\xi_1)\Psi(\xi_2)\ket=\frac{\hbar\del_0A(z_1)-\hbar\del_0A(z_2)}{\xi_1-\xi_2}
e^{A(z_1)+A(z_2)}.
\end{aligned}
\end{equation}
At the leading order in the genus expansion, the prefactor becomes
\begin{equation}
\begin{aligned}
 \frac{\hbar}{\gs}\frac{\del_0A_0(z_1)-\del_0A_0(z_2)}{\xi_1-\xi_2}&=\rt{2}\frac{\til{z}_1
-\til{z}_2}{z_1^2-z_2^2}=\frac{\rt{2}}{\til{z}_1+\til{z}_2},
\end{aligned} 
\label{eq:til-ann}
\end{equation} 
where we used the relation obtained from \eqref{eq:Veff}
\begin{equation}
\begin{aligned}
 \del_0A_0(z)=\til{z}.
\end{aligned} 
\end{equation}
The difference between the prefactors
of \eqref{eq:z-ann} and \eqref{eq:til-ann}
is compensated by
\begin{equation}
\begin{aligned}
 &\Bigl[A(z_1,z_2)
 -A(z_1)-A(z_2)\Bigr]_{\cO(\gs^0)}\\
=&\frac{1}{2}\left[D(z_1)+D(z_2)\right]^2 F_0
     -\frac{1}{2}D(z_1)^2 F_0-\frac{1}{2}D(z_2)^2 F_0\\
=&D(z_1)D(z_2)F_0\\
=&\int_0^{u_0}\frac{du}{\rt{(z^2_1-2u)(z_2^2-2u)}}\\
=&\log\left(\frac{z_1+z_2}{\til{z}_1+\til{z}_2}\right).
\end{aligned} 
\end{equation}
Thus the two expressions \eqref{eq:z-ann} and \eqref{eq:til-ann} 
are consistent at this order. 
\subsection{FZZT amplitude and CD kernel}
As shown in \cite{Strahov:2002zu}, the correlator of $2k$ determinants
in the finite $N$ matrix model is written in terms of the CD kernel
as
\begin{equation}
\begin{aligned}
 \Bigl\bra\prod_{i=1}^k\Psi_N(\xi_i)\Psi_N(\eta_i)\Bigr\ket
&=\frac{\det\bigl(K_{N+k}(\xi_i,\eta_j)\bigr)}{\lap(\xi)\lap(\eta)}\\
&=\frac{1}{\lap(\xi)\lap(\eta)}
\left|
\begin{matrix}
 K_{N+k}(\xi_1,\eta_1)&\cdots &K_{N+k}(\xi_1,\eta_k)\\
K_{N+k}(\xi_2,\eta_1)&\cdots &K_{N+k}(\xi_2,\eta_k)\\
\vdots &\cdots &\vdots\\
K_{N+k}(\xi_k,\eta_1)&\cdots& K_{N+k}(\xi_k,\eta_k)
\end{matrix}
\right|.
\end{aligned} 
\label{eq:even-det}
\end{equation}
For the odd number of determinants, the correlator can be obtained by sending $\eta_k\to\infty$ in \eqref{eq:even-det}
\begin{equation}
\begin{aligned}
 \Bigl\bra\prod_{i=1}^k\Psi_N(\xi_i)\prod_{j=1}^{k-1}\Psi_N(\eta_j)\Bigr\ket
&=\frac{\det\bigl(K_{N+k}(\xi_i,\eta_j)\big|\psi_{N+k-1}(\xi_i)\bigr)}{\lap(\xi)\lap(\eta)}\\
&=\frac{1}{\lap(\xi)\lap(\eta)}
\left|
\begin{matrix}
 K_{N+k}(\xi_1,\eta_1)&\cdots &K_{N+k}(\xi_1,\eta_{k-1})&\psi_{N+k-1}(\xi_1)\\
K_{N+k}(\xi_2,\eta_1)&\cdots &K_{N+k}(\xi_2,\eta_{k-1})&\psi_{N+k-1}(\xi_2)\\
\vdots &\cdots&\vdots &\vdots\\
K_{N+k}(\xi_k,\eta_1)&\cdots& K_{N+k}(\xi_k,\eta_{k-1})&\psi_{N+k-1}(\xi_k)
\end{matrix}
\right|.
\end{aligned} 
\label{eq:odd-det}
\end{equation}
In the double scaling limit, we find
\begin{equation}
\begin{aligned}
 \Bigl\bra\prod_{i=1}^k\Psi(\xi_i)\Psi(\eta_i)\Bigr\ket
&=\frac{\det\bigl(K(\xi_i,\eta_j)\bigr)}{\lap(\xi)\lap(\eta)},\\
\Bigl\bra\prod_{i=1}^k\Psi(\xi_i)\prod_{j=1}^{k-1}\Psi(\eta_j)\Bigr\ket
&=\frac{\det\bigl(K(\xi_i,\eta_j)\big|\psi(\xi_i)\bigr)}{\lap(\xi)\lap(\eta)},
\end{aligned} 
\label{eq:corr-CD-double}
\end{equation}
where $\psi(\xi)$ and $K(\xi,\eta)$ are the BA function and the CD kernel 
\eqref{eq:CDkernel}, 
respectively. In other words, the correlator of the even number of FZZT branes can be 
written as a product of two-point functions $K(\xi,\eta)$.
Although the permutation symmetry of the parameters $\xi_i$ and $\eta_i$ is not manifest
in \eqref{eq:corr-CD-double}, the result is symmetric as proved in \cite{Strahov:2002zu}.
Of course, one can use \eqref{eq:lap-d} for the multi-point function of FZZT branes.
However, $\del_0^n\psi(\xi)$ with $n\geq2$
can be reduced to a linear combination of 
$\psi(\xi)$ and $\del_0\psi(\xi)$ using the Schr\"{o}dinger equation
\eqref{eq:Sch},
and after this rewriting
one finds that \eqref{eq:lap-d} is identical with \eqref{eq:corr-CD-double}.

\section{Correlator of FZZT branes and macroscopic loops}\label{sec:corr}
In this section, we consider the correlator of FZZT branes $\Psi(\xi_i)$ 
and macroscopic loop operators $Z(\bt_j)$. 

\subsection{Bra--ket notation}

To describe our results,
it is convenient to introduce the bra--ket notation as follows.

Let $\hat{x}$ be the coordinate operator
and $\hat{\partial}_x$ be its conjugate ``momentum operator''
satisfying
\begin{align}
{}[\hat{\partial}_x,\hat{x}]=1.
\end{align}
We then introduce the operator
\begin{align}
Q:=\hbar^2\hat{\partial}_x^2+\hat{u},\qquad
\hat{u}:=u\Big|_{t_0=\hat{x}},
\end{align}
where $u$ is the specific heat \eqref{eq:udef}.
We decided
not to put a hat $\hat{}\,$ on $Q$ just for notational simplicity.

For a Roman letter $x$, we let $|x\ket$ denote
the coordinate eigenstate, e.g.
\begin{align}
\hat{x}|x\ket = x|x\ket,\qquad
\hat{x}|t_0\ket = t_0|t_0\ket.
\end{align}
For a Greek letter $\xi$, on the other hand, we let $|\xi\ket$ denote
an eigenstate of $Q$ with eigenvalue $\xi$:
\begin{align}
Q|\xi\ket = \xi|\xi\ket.
\label{eq:Qket}
\end{align}
These states are normalized such that
\begin{align}
1=\int_{-\infty}^\infty d\xi |\xi\ket\bra\xi|,\qquad
\psi(\xi;x)=\bra x|\xi\ket=\bra\xi|x\ket,
\label{eq:bknorm}
\end{align}
where $\psi(\xi;x)$ is the BA function.
Another important element is the projection operator
\begin{align}
\Pi:=\frac{1}{\hbar}\int_{-\infty}^{t_0}dx|x\ket\bra x|.
\end{align}
As discussed in \cite{Banks:1989df}, $\Pi$ 
can be interpreted as the projection below the Fermi level $t_0$.
In terms of $\Pi$ the CD kernel \eqref{eq:CDkernel}
is expressed as
\begin{align}
K(\xi,\eta)=\bra\xi|\Pi|\eta\ket.
\end{align}
Note that $\Pi$ satisfies
\begin{align}
\hbar\partial_0\Pi = |t_0\ket\bra t_0|.
\end{align}

Note also that $\bra t_0|$ and $|t_0\ket$ satisfy
\begin{align}
\bra t_0|Q=(\hbar^2\partial_0^2+u)\bra t_0|,\qquad
Q|t_0\ket =(\hbar^2\partial_0^2+u)|t_0\ket,
\end{align}
so that the Schr\"odinger equation \eqref{eq:Sch} holds.

\subsection{Correlator of one FZZT brane and macroscopic loops}
As a first step, we consider the correlator of one FZZT brane and macroscopic loops.
Following \cite{Blommaert:2019wfy},
let us consider the correlator of $\det(\xi-M)$ and $\Tr e^{\bt M}$ at finite $N$:
\begin{equation}
\begin{aligned}
 &\big\bra\Tr e^{\bt M}\Psi_N(\xi)\big\ket\\
=&\frac{1}{N!Z_N}
\int \prod_{i=1}^N d\la_ie^{-V(\la_i)}\lap(\la)^2\sum_{i=1}^N e^{\bt\la_i}
\frac{e^{-\hf V(\xi)}}{\rt{h_{N-1}}}\prod_{i=1}^N(\xi-\la_i)\\
=&\frac{1}{(N-1)!Z_{N-1}}\\
&\times
\int d\la e^{\bt\la}(\xi-\la)
\int \prod_{i=1}^{N-1} d\la_i'e^{-V(\la_i')}\lap(\la')^2\frac{e^{-V(\la)}}{h_{N-1}}
\frac{e^{-\hf V(\xi)}}{\rt{h_{N-1}}}\prod_{i=1}^{N-1}(\la-\la_i')^2(\xi-\la_i')\\
=&(r_{N-1}^3)^{-\frac{1}{2}}\int d\la e^{\bt\la}(\xi-\la)\big\bra\Psi_{N-1}(\la)^2\Psi_{N-1}(\xi)\big\ket,
\end{aligned} 
\end{equation}
where $Z_N$ is given by \eqref{eq:ZN-norm}
and $M$ is related to $H$ by $M=-H$ (see footnote \ref{foot:M}).
It follows from
\eqref{eq:corr-CD-double} that
in the double scaling limit this becomes
\begin{equation}
\begin{aligned}
 \bra Z(\bt)\Psi(\xi)\ket&=\int d\la e^{\bt\la}
(\xi-\la)\frac{1}{\xi-\la}\left|
\begin{matrix}
 K(\la,\la)&\psi(\la)\\
K(\xi,\la)&\psi(\xi)
\end{matrix}
\right|\\
&=\int d\la e^{\bt\la}\Bigl[K(\la,\la)\psi(\xi)-K(\xi,\la)\psi(\la)\Bigr].
\end{aligned} 
\label{eq:Z-FZZT-full}
\end{equation}
In appendix~\ref{app:inverse}, we will give an alternative derivation of this result.
Note that $\rho(\la)=K(\la,\la)$ is the eigenvalue density. Thus the first term 
of \eqref{eq:Z-FZZT-full} is
the disconnected part $\bra Z(\bt)\ket\bra\Psi(\xi)\ket$ and the second term is the connected part
\begin{equation}
\begin{aligned}
 \bra Z(\bt)\Psi(\xi)\ket_\conn&=-\int d\la e^{\bt\la}
\psi(\la)K(\la,\xi)\\
&=-\int d\la e^{\bt\la}
\bra t_0|\la\ket\bra\la|\Pi|\xi\ket\\
&=-\bra t_0|e^{\bt Q}\Pi |\xi\ket.
\end{aligned} 
\label{eq:Z-FZZT-conn}
\end{equation}
In the last step we have used \eqref{eq:Qket} and \eqref{eq:bknorm}.

Next consider the correlator of one FZZT brane with two macroscopic loops
\begin{equation}
\begin{aligned}
 &\big\bra \Tr e^{\bt_1M}\Tr e^{\bt_2M}\Psi_N(\xi)\big\ket\\
=&\frac{1}{N!Z_N}
\int \prod_{i=1}^N d\la_ie^{-V(\la_i)}\lap(\la)^2\sum_{i,j=1}^N e^{\bt_1\la_i+\bt_2\la_j}
\frac{e^{-\hf V(\xi)}}{\rt{h_{N-1}}}\prod_{i=1}^N(\xi-\la_i)\\
=&(r_{N-1}^3)^{-\frac{1}{2}}\int d\la e^{(\bt_1+\bt_2)\la}(\xi-\la)\big\bra\Psi_{N-1}(\la)^2\Psi_{N-1}(\xi)\big\ket\\
+&(r_{N-1}^3r_{N-2}^5)^{-\frac{1}{2}}\int \prod_{i=1,2}d\la_i e^{\bt_i\la_i}(\xi-\la_i)(\la_1-\la_2)^2
\big\bra\Psi_{N-2}(\la_1)^2\Psi_{N-2}(\la_2)^2\Psi_{N-2}(\xi)\big\ket.
\end{aligned} 
\end{equation}
From \eqref{eq:corr-CD-double}, in the double scaling limit this becomes
\begin{equation}
\begin{aligned}
 \big\bra Z(\bt_1)Z(\bt_2)\Psi(\xi)\big\ket&=
\big\bra Z(\bt_1+\bt_2)\Psi(\xi)\big\ket\\
&+\int \prod_{i=1,2}d\la_i e^{\bt_i\la_i}
\left|
\begin{matrix}
 K(\la_1,\la_1)&K(\la_1,\la_2) &  \psi(\la_1)\\
K(\la_2,\la_1) & K(\la_2,\la_2)&  \psi(\la_2)\\
K(\xi,\la_1) & K(\xi,\la_2)  &\psi(\xi)
\end{matrix}
\right|\\
&=\big\bra Z(\bt_1)Z(\bt_2)\big\ket\psi(\xi)-\bra t_0|e^{(\bt_1+\bt_2)Q}\Pi |\xi\ket\\
&\quad -\big\bra Z(\bt_1)\big\ket \bra t_0|e^{\bt_2Q}\Pi|\xi\ket-\big\bra Z(\bt_2)\big\ket \bra t_0|e^{\bt_1Q}\Pi|\xi\ket\\
&\quad +\bra t_0|e^{\bt_1Q}\Pi e^{\bt_2Q}\Pi|\xi\ket+\bra t_0|e^{\bt_2Q}\Pi e^{\bt_1Q}\Pi|\xi\ket.
\end{aligned} 
\end{equation}
One can see that the connected part is given by
\begin{equation}
\begin{aligned}
 \big\bra Z(\bt_1)Z(\bt_2)\Psi(\xi)\big\ket_\conn=&-\bra t_0|e^{(\bt_1+\bt_2)Q}\Pi |\xi\ket\\
&+\bra t_0|e^{\bt_1Q}\Pi e^{\bt_2Q}\Pi|\xi\ket+\bra t_0|e^{\bt_2Q}\Pi e^{\bt_1Q}\Pi|\xi\ket.
\end{aligned} 
\label{eq:ZZ-FZZT-conn}
\end{equation}

In the same way, one can calculate the correlator
of one FZZT brane and three macroscopic loops. The result is
\begin{align}
\begin{aligned}
&\hspace{-1em}\big\bra Z(\bt_1)Z(\bt_2)Z(\bt_3)\Psi(\xi)\big\ket_\conn\\[1ex]
&=-\,\bbra{t_0}e^{(\beta_1+\beta_2+\beta_3)Q}\Pi\kket{\xi}\\[1ex]
&\hspace{1em}
  +\sum_{\substack{(i,j,k)\,=\,\mbox{\scriptsize cyclic}\\
                   \mbox{\scriptsize permutations of $(1,2,3)$}}}\left[
   \bbra{t_0}e^{(\beta_i+\beta_j)Q}\Pi e^{\beta_k Q}\Pi\kket{\xi}
  +\bbra{t_0}e^{\beta_k Q}\Pi e^{(\beta_i+\beta_j)Q}\Pi\kket{\xi}\right]\\[1ex]
&\hspace{1em}
  -\sum_{\substack{(i,j,k)\,=\,\mbox{\scriptsize all possible}\\
                   \mbox{\scriptsize permutations of $(1,2,3)$}}}
 \bbra{t_0}e^{\beta_i Q}\Pi e^{\beta_j Q}\Pi e^{\beta_k Q}\Pi\kket{\xi}.
\end{aligned}
\label{eq:ZZZ-FZZT-conn}
\end{align}

From \eqref{eq:Z-FZZT-conn}, \eqref{eq:ZZ-FZZT-conn}
and \eqref{eq:ZZZ-FZZT-conn}
it is natural to conjecture that the generating function of the connected part
of the correlator of one FZZT brane and arbitrary number of macroscopic loops
is
\begin{equation}
\begin{aligned}
 \big\bra Z(\bt_1)\cdots Z(\bt_n)\Psi(\xi)\big\ket_\conn
=\til{\psi}(\xi)\Big|_{\cO(w_1\cdots w_n)},
\end{aligned} 
\label{eq:Z-FZZT-gen}
\end{equation}
where
\begin{equation}
\begin{aligned}
\til{\psi}(\xi)=\bra t_0|G^{-1}|\xi\ket
\end{aligned} 
\label{eq:til-psi}
\end{equation}
with
\begin{equation}
\begin{aligned}
 G&=1+A\Pi,\quad A=-1+\prod_{i=1}^n(1+w_ie^{\bt_iQ}). 
\end{aligned} 
\label{eq:GA-def}
\end{equation}

As a non-trivial check of this conjecture,
in what follows we will prove that $\til{\psi}(\xi)$
given above obeys the Schr\"odinger equation.
We first notice that the connected correlator
in \eqref{eq:Z-FZZT-gen} is obtained by applying
the boundary creation operators
to
the BA function $\psi(\xi)$
\begin{equation}
\begin{aligned}
 \big\bra Z(\bt_1)\cdots Z(\bt_n)\Psi(\xi)\big\ket_\conn=B(\bt_1)\cdots B(\bt_n)\psi(\xi)=:\psi_n(\beta_1,\ldots,\beta_n;\xi).
\end{aligned} 
\end{equation}
Applying
the boundary creation operator
to
the both sides of the Schr\"{o}dinger equation
\eqref{eq:Sch}, we obtain
\begin{align}
(\hbar^2\partial_0^2+u-\xi)\psi_n(\beta_1,\ldots,\beta_n;\xi)
 +2\hbar^2\sum_{I\subsetneq S}\partial_0^2 Z_{|S-I|}\psi_{|I|}=0.
\label{eq:psindiffeq}
\end{align}
Here
\begin{align}
\begin{aligned}
Z_{|I|}
 &=\big\bra Z(\bt_{i_1})\cdots Z(\bt_{i_I})\big\ket_\conn,\quad
\psi_{|I|}
 =\psi_{|I|}(\beta_{i_1},\ldots,\beta_{i_I};\xi),\quad
\psi_0=\psi(\xi)
\end{aligned}
\end{align}
with $I=\{i_1,i_2,\ldots,i_{|I|}\}$, $S=\{1,2,\ldots,n\}$
and the sum is taken for all possible proper subsets $I$ of $S$
including the empty set.
In terms of the generating function,
\eqref{eq:psindiffeq} is simply written as
\begin{equation}
\begin{aligned}
(\hbar^2\partial_0^2+u-\xi)
\til{\psi}(\xi)+2\hbar^2\del_0^2\cZ\til{\psi}(\xi)=0.
\end{aligned} 
\label{eq:psi-Sch}
\end{equation}
Here $\cZ$ is the generating function of the connected correlator 
of macroscopic loops such that
\begin{equation}
\begin{aligned}
 \big\bra Z(\bt_1)\cdots Z(\bt_n)\big\ket_\conn
=\cZ\Big|_{\cO(w_1\cdots w_n)}.
\end{aligned} 
\label{eq:Z-gen}
\end{equation}
As shown in \cite{Banks:1989df,Okuyama:2018aij,Okuyama:2019xbv} it is given by
\begin{equation}
\begin{aligned}
 \cZ=\Tr\log G.
\end{aligned} 
\end{equation}
Thus \eqref{eq:psi-Sch} serves as the Schr\"odinger equation
for $\til{\psi}(\xi)$. 
We will check if our conjectural expression \eqref{eq:til-psi}
of $\til{\psi}(\xi)$
indeed satisfies \eqref{eq:psi-Sch}. To do this, let us
first compute $\del_0^2 \cZ$. Using 
\begin{equation}
\begin{aligned}
 \hbar\del_0\Pi=|t_0\ket\bra t_0|,\quad
\hbar\del_0G^{-1}=-G^{-1}A|t_0\ket\bra t_0|G^{-1},
\end{aligned} 
\end{equation}
we find
\begin{equation}
\begin{aligned}
 \hbar^2\del_0^2\cZ&=\hbar^2\del_0\Tr G^{-1}A\del_0\Pi\\
&=-\hbar^2\Tr G^{-1}A\del_0\Pi G^{-1} A\del_0\Pi+\hbar^2\Tr G^{-1}A\del_0^2\Pi\\
&=-\bra t_0|G^{-1}A|t_0\ket^2+\hbar\bra \del_0 t_0|G^{-1}A|t_0\ket +
\hbar\bra t_0|G^{-1}A|\del_0 t_0\ket,
\end{aligned} 
\end{equation}
where we have introduced the notation
\begin{align}
\bra\partial_0t_0|:=\partial_0\bra t_0|,\qquad
|\partial_0t_0\ket:=\partial_0|t_0\ket.
\end{align}
Next consider the first term of \eqref{eq:psi-Sch}. It is rewritten as
\begin{equation}
\begin{aligned}
(\hbar^2\partial_0^2+u-\xi)\til{\psi}(\xi)
&=\bra t_0|QG^{-1}|\xi\ket-\bra t_0|G^{-1}\xi|\xi\ket
+2\hbar^2\bra\del_0 t_0|\del_0 G^{-1}|\xi\ket+\hbar^2\bra t_0|\del_0^2G^{-1}|\xi\ket\\
&=\bra t_0|[Q,G^{-1}]|\xi\ket+2\hbar^2\bra\del_0 t_0|\del_0 G^{-1}|\xi\ket+\hbar^2\bra t_0|\del_0^2G^{-1}|\xi\ket.
\end{aligned} 
\label{eq:Q-xi}
\end{equation}
The first term of \eqref{eq:Q-xi} is
\begin{equation}
\begin{aligned}
\bra t_0| [Q,G^{-1}]|\xi\ket&=-\bra t_0|G^{-1}A[Q,\Pi]G^{-1}|\xi\ket \\
&=-\hbar\bra t_0|G^{-1}A\Bigl[|\del_0 t_0\ket \bra t_0|-| t_0\ket \bra\del_0 t_0|\Bigr]G^{-1}|\xi\ket\\
&=-\hbar\bra t_0|G^{-1}A|\del_0 t_0\ket \til{\psi}(\xi)+
\hbar\bra t_0|G^{-1}A| t_0\ket\bra\del_0 t_0|G^{-1}|\xi\ket.
\end{aligned} 
\end{equation}
The second term of \eqref{eq:Q-xi} is
\begin{equation}
\begin{aligned}
 -2\hbar^2\bra\del_0t_0|G^{-1}A\del_0\Pi G^{-1}|\xi\ket=
-2\hbar\bra\del_0t_0|G^{-1}A|t_0\ket\til{\psi}(\xi).
\end{aligned} 
\end{equation}
The third term of \eqref{eq:Q-xi} is
\begin{equation}
\begin{aligned}
&\hspace{-1em}
 -\hbar^2\bra t_0|\del_0\Bigl(G^{-1}A\del_0\Pi G^{-1}\Bigr)|\xi\ket\\
&=
2\hbar^2\bra t_0|G^{-1}A\del_0\Pi G^{-1}A\del_0\Pi G^{-1}|\xi\ket-
\hbar^2\bra t_0|G^{-1}A\del_0^2\Pi G^{-1}|\xi\ket\\
&=2\bra t_0|G^{-1}A|t_0\ket^2\til{\psi}(\xi)-\hbar\bra t_0|G^{-1}A|\del_0 t_0\ket\til{\psi}(\xi)
-\hbar\bra t_0|G^{-1}A|t_0\ket\bra\del_0 t_0|G^{-1}|\xi\ket.
\end{aligned} 
\end{equation}
Finally we find
\begin{equation}
\begin{aligned}
(\hbar^2\partial_0^2+u-\xi)\til{\psi}(\xi)
&=2\Bigl(\bra t_0|G^{-1}A|t_0\ket^2
-\hbar\bra \del_0t_0|G^{-1}A|t_0\ket-\hbar\bra t_0|G^{-1}A|\del_0 t_0\ket\Bigr)\til{\psi}(\xi)\\
&=-2\hbar^2\del_0^2\cZ \til{\psi}(\xi).
\end{aligned} 
\end{equation}
Thus $\til{\psi}(\xi)$ in \eqref{eq:til-psi}
satisfies the necessary equation \eqref{eq:psi-Sch},
and we conclude that $\til{\psi}(\xi)$ is the generating function 
of the connected correlators \eqref{eq:Z-FZZT-gen}.

\subsection{General correlator of FZZT branes and macroscopic loops}
We found that the generating function \eqref{eq:til-psi} 
satisfies \eqref{eq:psi-Sch}.
The key observation is that \eqref{eq:psi-Sch} is itself a
Schr\"{o}dinger equation with modified potential
\begin{equation}
\begin{aligned}
 (\hbar^2\del_0^2+\til{u})\til{\psi}(\xi)=\xi \til{\psi}(\xi),
\end{aligned} 
\end{equation}
where $\til{u}=2\hbar^2\del_0^2\til{F}$ with $\til{F}$ being
\begin{equation}
\begin{aligned}
\til{F}&=F+\cZ=F+\Tr\log G.
\end{aligned} 
\end{equation}
This suggests that $\til{\psi}(\xi)$ is a BA function in a certain
modified background.

To see this, let us consider
the correlator of $Z(\bt)$'s and FZZT branes at finite $N$
\begin{equation}
\begin{aligned}
 &\Biggl\bra\prod_{i=1}^nZ(\bt_i)\prod_{j=1}^k\Psi_N(\xi_j)
\Psi_N(\eta_j)\Biggr\ket\\
=&
\Biggl\bra e^{\sum_{i=1}^nw_iZ(\bt_i)}\prod_{j=1}^k\Psi_N(\xi_j)
\Psi_N(\eta_j)\Biggr\ket\Bigg|_{\cO(w_1\cdots w_n)}\\
=&\frac{1}{Z_N}\int dM e^{-\Tr \til{V}(M)}
\prod_{j=1}^k\Psi_N(\xi_j)
\Psi_N(\eta_j)\Bigg|_{\cO(w_1\cdots w_n)},
\end{aligned} 
\label{eq:deformedV}
\end{equation}
where the deformed potential $\til{V}(M)$ is given by
\begin{equation}
\begin{aligned}
 \til{V}(M)=V(M)-\sum_{i=1}^n w_i e^{\bt_i M}.
\end{aligned} 
\end{equation}
Since \eqref{eq:deformedV} is just the correlator of determinants in the deformed
matrix model integral, it also takes the form \eqref{eq:even-det}.
In the double scaling limit we find
\begin{equation}
\begin{aligned}
  \Biggl\bra\prod_{i=1}^nZ(\bt_i)\prod_{j=1}^k\Psi(\xi_j)\Psi(\eta_j)\Biggr\ket
=
\det G\frac{\det\bigl(\til{K}(\xi_i,\eta_j)\bigr)}{\lap(\xi)\lap(\eta)}\Bigg|_{\cO(w_1\cdots w_n)}.
\end{aligned} 
\label{eq:even-gen}
\end{equation}
The overall
factor
$\det G=e^{\til{F}-F}$ accounts
for the different normalization
of the matrix integrals. 
For odd number of FZZT branes we have
\begin{equation}
\begin{aligned}
  \Biggl\bra\prod_{i=1}^nZ(\bt_i)\prod_{j=1}^k\Psi(\xi_j)
\prod_{l=1}^{k-1}\Psi(\eta_l)\Biggr\ket
=
\det G\frac{\det\bigl(\til{K}(\xi_i,\eta_j)\big|\til{\psi}(\xi_i)\bigr)}{\lap(\xi)\lap(\eta)}\Bigg|_{\cO(w_1\cdots w_n)}.
\end{aligned} 
\label{eq:odd-gen}
\end{equation}

The CD kernel in the modified potential
can be found from the condition \eqref{eq:del-K}
\begin{equation}
\begin{aligned}
 \hbar\del_0\til{K}(\xi,\eta)=\til{\psi}(\xi)\til{\psi}(\eta).
\end{aligned} 
\label{eq:del-tilK}
\end{equation}
We find that $\til{K}(\xi,\eta)$ is given by
\begin{equation}
\begin{aligned}
 \til{K}(\xi,\eta)=\bra\eta|\Pi G^{-1}|\xi\ket.
\end{aligned} 
\label{eq:til-K}
\end{equation}
Let us see that \eqref{eq:til-K} indeed satisfies \eqref{eq:del-tilK}:
\begin{equation}
\begin{aligned}
 \hbar\del_0\til{K}(\xi,\eta)&=\hbar \bra\eta|\del_0\Pi G^{-1}|\xi\ket
-\hbar\bra\eta|\Pi G^{-1}A\del_0\Pi G^{-1}|\xi\ket\\
&=\bra\eta|t_0\ket\bra t_0| G^{-1}|\xi\ket
-\bra\eta|\Pi G^{-1}A|t_0\ket\bra t_0| G^{-1}|\xi\ket\\
&=\bra\eta|(1-\Pi G^{-1}A)|t_0\ket\bra t_0| G^{-1}|\xi\ket.
\end{aligned}
\label{eq:dtilK} 
\end{equation}
The combination $1-\Pi G^{-1}A$ in the first factor is written as
\begin{equation}
\begin{aligned}
 1-\Pi G^{-1}A&=1-\Pi(1-A\Pi+A\Pi A\Pi+\cdots)A\\
&=1-\Pi A+\Pi A\Pi A-\Pi A\Pi A\Pi A+\cdots\\
&=(1+\Pi A)^{-1}={}^t G^{-1}. 
\end{aligned} 
\end{equation} 
Thus \eqref{eq:dtilK} becomes
\begin{equation}
\begin{aligned}
 \hbar\del_0\til{K}(\xi,\eta)=\bra\eta|{}^tG^{-1}|t_0\ket\bra t_0|G^{-1}|\xi\ket=
\til{\psi}(\xi)\til{\psi}(\eta).
\end{aligned} 
\end{equation}
This confirms \eqref{eq:til-K}.

To summarize, the generating function of the correlator
of FZZT branes and macroscopic loops is given by
\eqref{eq:even-gen} for even number of FZZT branes and 
\eqref{eq:odd-gen} for odd number of FZZT branes.
Our formulae \eqref{eq:even-gen} and \eqref{eq:odd-gen}
do not rely on the genus expansion and in principle
they can be defined non-perturbatively.

\section{Airy case}\label{sec:airy}
In this section we consider the Airy case corresponding to
the trivial background $t_k=0~(k\geq1)$.
In this case $u=t_0$ and the Schr\"{o}dinger equation for the BA
function is
\begin{equation}
\begin{aligned}
 (\hbar^2\del_0^2+t_0)\psi(\xi)=\xi\psi(\xi).
\end{aligned} 
\end{equation}
The solution to this equation is given by the Airy function
\begin{equation}
\begin{aligned}
 \psi(\xi)=\bra t_0|\xi\ket=\hbar^{-\frac{2}{3}}\text{Ai}\Bigl[\hbar^{-\frac{2}{3}}(\xi-t_0)\Bigr].
\end{aligned} 
\end{equation}
\subsection{$Z(\bt)$-FZZT amplitude}
Let us apply our formula \eqref{eq:Z-FZZT-conn} to the Airy case. 
For simplicity we set $t_0=0$.
Then \eqref{eq:Z-FZZT-conn} becomes
\begin{equation}
\begin{aligned}
 \bra Z(\bt)\Psi(\xi)\ket_\conn&=-\bra 0|e^{\bt Q}\Pi|\xi\ket
=-\int_{-\infty}^0\frac{dx}{\hbar}\bra 0|e^{\bt Q}|x\ket\bra x|\xi\ket.
\end{aligned} 
\end{equation}
where $\bra0|=\bra x=0|$.
As shown in \cite{okounkov2002generating} the matrix element of $e^{\bt Q}$
is given by
\begin{equation}
\begin{aligned}
 \bra x_1|e^{\bt Q}|x_2\ket=\frac{1}{2\rt{\pi\bt}}
\exp\left[\frac{\bt^3\hbar^2}{12}+\frac{\bt}{2}(x_1+x_2)-\frac{(x_1-x_2)^2}{4\bt\hbar^2}\right].
\end{aligned}
\label{eq:eQ-ele} 
\end{equation}
Thus we find
\begin{equation}
\begin{aligned}
 \bra Z(\bt)\Psi(\xi)\ket_\conn&=-
\frac{e^{\frac{\bt^3\hbar^2}{12}}}{2\hbar\rt{\pi\bt}}\int_{-\infty}^0 dx
e^{\hf\bt x-\frac{x^2}{4\bt\hbar^2}}
\bra x|\xi\ket.
\end{aligned} 
\label{eq:airy-zf}
\end{equation}
Using the integral representation of the Airy function
\begin{equation}
\begin{aligned}
 \bra x|\xi\ket=\int_{-\infty}^\infty\frac{d\mu}{2\pi\hbar}
e^{\frac{\ri}{\hbar}\bigl[\frac{\mu^3}{3}+(\xi-x)\mu\bigr]},
\end{aligned} 
\end{equation}
\eqref{eq:airy-zf} becomes
\begin{equation}
\begin{aligned}
\bra Z(\bt)\Psi(\xi)\ket_\conn=-\hf
e^{\frac{\bt^3\hbar^2}{12}}\int_{-\infty}^\infty\frac{d\mu}{2\pi\hbar}
e^{\frac{\ri}{\hbar}\bigl[\frac{\mu^3}{3}+\xi\mu\bigr]+\qu\bt(\bt\hbar-2\ri\mu)^2}
\text{Erfc}\Biggl(\hf\rt{\bt}(\bt\hbar-2\ri\mu)\Biggr).
\end{aligned} 
\end{equation}
In the $\hbar\to0$ limit,
the $\mu$ integral can be evaluated by the saddle point approximation.
The saddle point is given by
\begin{equation}
\begin{aligned}
 \mu_*^2=-\xi=-\frac{z^2}{2}.
\end{aligned} 
\end{equation}
Thus in the leading order approximation of $\hbar$ expansion we find
\begin{equation}
\begin{aligned}
 \bra Z(\bt)\Psi(\xi)\ket_\conn\approx -\psi(\xi)\times \hf e^{\hf\bt z^2}
\text{Erfc}\Biggl(\rt{\frac{\bt z^2}{2}}\Biggr).
\end{aligned} 
\end{equation}
This reproduces the ``half-wormhole'' amplitude \eqref{eq:z-HW}, as expected.

Let us consider the eigenvalue density deformed by the FZZT brane
\begin{equation}
\begin{aligned}
 \frac{\bra Z(\bt)\Psi(\xi)\ket}{\bra\Psi(\xi)\ket}=\int d\la e^{\bt\la}\rho_1(\la,\xi).
\end{aligned} 
\end{equation}
Here we consider the full correlator, not the connected part 
$\bra Z(\bt)\Psi(\xi)\ket_\conn$. From \eqref{eq:Z-FZZT-full} we find
\begin{equation}
\begin{aligned}
 \rho_1(\la,\xi)=K(\la,\la)-\frac{\psi(\la)}{\psi(\xi)}K(\la,\xi).
\end{aligned} 
\label{eq:rho-la-xi}
\end{equation}
In the Airy case we have
\begin{equation}
\begin{aligned}
\psi(\la)&=\text{Ai}(\la),\\
 K(\la,\xi)&=\frac{\text{Ai}(\la)\text{Ai}'(\xi)
-\text{Ai}'(\la)\text{Ai}(\xi)}{\la-\xi},\\
K(\la,\la)&=\text{Ai}'(\la)^2-\text{Ai}''(\la)\text{Ai}(\la),
\end{aligned} 
\label{eq:CD-Airy}
\end{equation}
where we have set $\hbar=1$ for simplicity.
Note that $\rho_1(\la,\xi)$ vanishes at $\la=\xi$
\begin{equation}
\begin{aligned}
 \lim_{\la\to\xi}\rho_1(\la,\xi)=0.
\end{aligned} 
\end{equation}
This is understood from the eigenvalue repulsion due to the insertion of
FZZT brane at $\la=\xi$. The density $\rho_1(\la,\xi)$ in \eqref{eq:rho-la-xi}
is not positive definite since the single determinant 
$\det(\xi-M)$ can take both positive and negative values. As discussed in \cite{Blommaert:2019wfy},
we can define a positive definite eigenvalue density
deformed by the two FZZT branes $\det(\xi-M)^2$, 
which we will consider in the next subsection.

\subsection{$Z(\bt)$-$(\text{FZZT})^2$ amplitude}
\begin{figure}[htb]
\centering
\subcaptionbox{$E'=-3$ \label{sfig:Em3-2}}{\includegraphics
[width=0.45\linewidth]{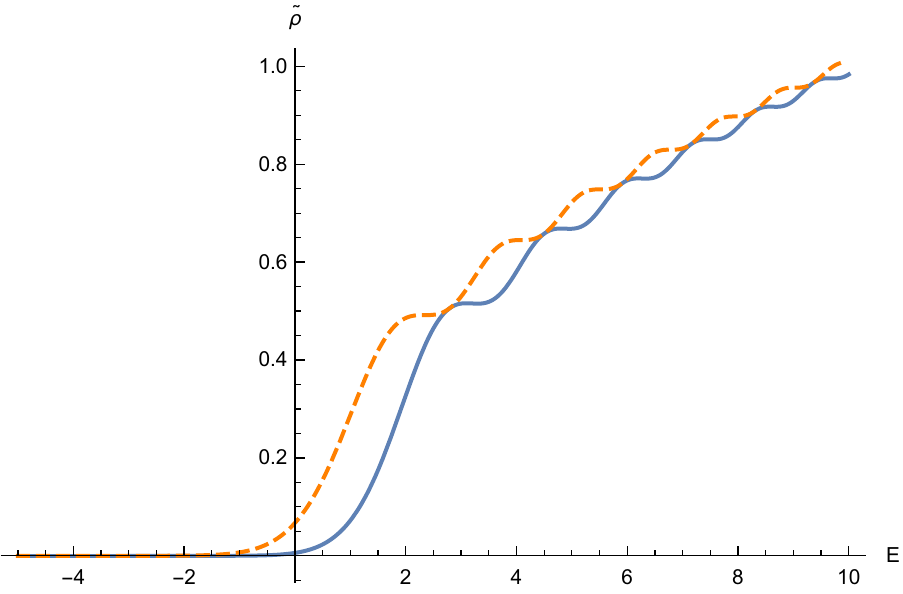}}
\hskip5mm
\subcaptionbox{$E'=6$ \label{sfig:E6-2}}{\includegraphics
[width=0.45\linewidth]{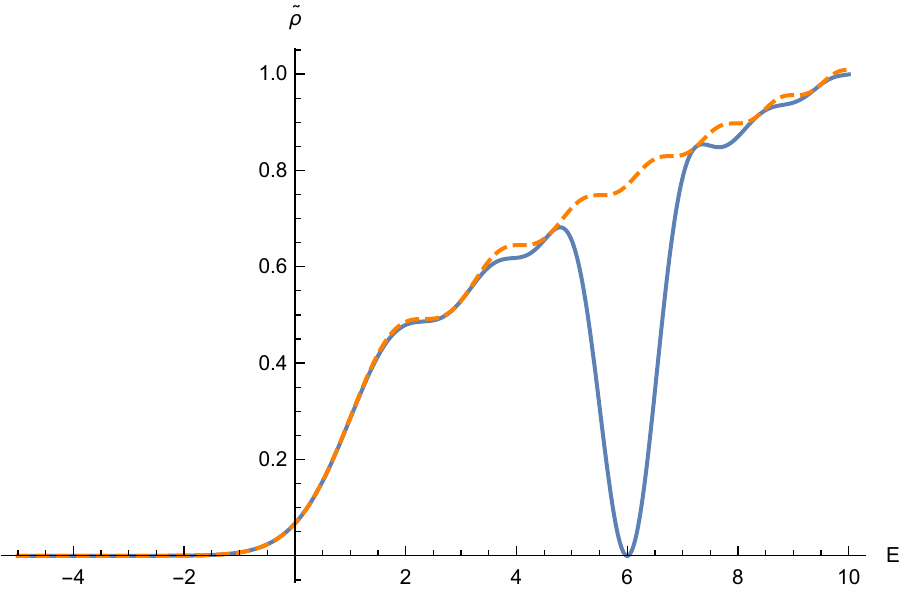}}
  \caption{
Plot 
of $\til{\rho}(E,E')$ in \eqref{eq:tilrho} 
for \subref{sfig:Em3-2} $E'=-3$ and \subref{sfig:E6-2} $E'=6$, 
as a function of $E$.
The solid curves are the deformed eigenvalue density $\til{\rho}(E,E')$ 
while
the orange dashed curve represents the original eigenvalue density
 $\til{\rho}(E)=K(-E,-E)$ without FZZT branes.
}
  \label{fig:til-rho2}
\end{figure}

Let us consider $Z(\bt)$-$(\text{FZZT})^2$ amplitude.
From our general formula \eqref{eq:even-gen} we find
\begin{equation}
\begin{aligned}
 \bra Z(\bt)\Psi(\xi)^2\ket&=\Tr (e^{\bt Q}\Pi)\bra\xi|\Pi|\xi\ket
-\bra\xi|\Pi e^{\bt Q}\Pi|\xi\ket\\
&=\int d\la e^{\bt\la}\Bigl[K(\la,\la)K(\xi,\xi)-K(\la,\xi)^2\Bigr].
\end{aligned}
\label{eq:Z-FF} 
\end{equation}
We define the deformed eigenvalue density due to the insertion of two
FZZT branes as
\begin{equation}
\begin{aligned}
 \frac{\bra Z(\bt)\Psi(\xi)^2\ket}{\bra\Psi(\xi)^2\ket}
=\int d\la e^{\bt\la}\rho_2(\la,\xi).
\end{aligned} 
\end{equation}
From \eqref{eq:Z-FF} and $\bra\Psi(\xi)^2\ket=K(\xi,\xi)$ we find
\begin{equation}
\begin{aligned}
 \rho_2(\la,\xi)
=K(\la,\la)-\frac{K(\la,\xi)^2}{K(\xi,\xi)}.
\end{aligned} 
\label{eq:rho2}
\end{equation}
We also define
\begin{equation}
\begin{aligned}
 \til{\rho}(E,E')=\rho_2(-E,-E').
\end{aligned} 
\label{eq:tilrho}
\end{equation}
Using the CD kernel for the Airy case \eqref{eq:CD-Airy}, we can evaluate
$\til{\rho}(E,E')$ numerically.
In Figure~\ref{fig:til-rho2}, we show 
the plot of $\til{\rho}(E,E')$ for
$E'=-3$ and $E'=6$.
When $E'>0$ we see a void near $E=E'$ (see Figure~\ref{sfig:E6-2}).
This reproduces the result of ``eigenbrane'' in \cite{Blommaert:2019wfy}.
On the other hand, when $E'<0$ the eigenvalues are pushed to the positive $E$ direction
due to the eigenvalue repulsion (see Figure~\ref{sfig:Em3-2}). This is qualitatively
similar to the result of inserting $K$ FZZT branes in the large $K$
't Hooft limit (see Figure~\ref{fig:rho}).

\subsection{Spectral form factor in FZZT brane background}
\begin{figure}[htb]
\centering
\subcaptionbox{$E'=-3$ \label{sfig:sff-em3}}{\includegraphics
[width=0.3\linewidth]{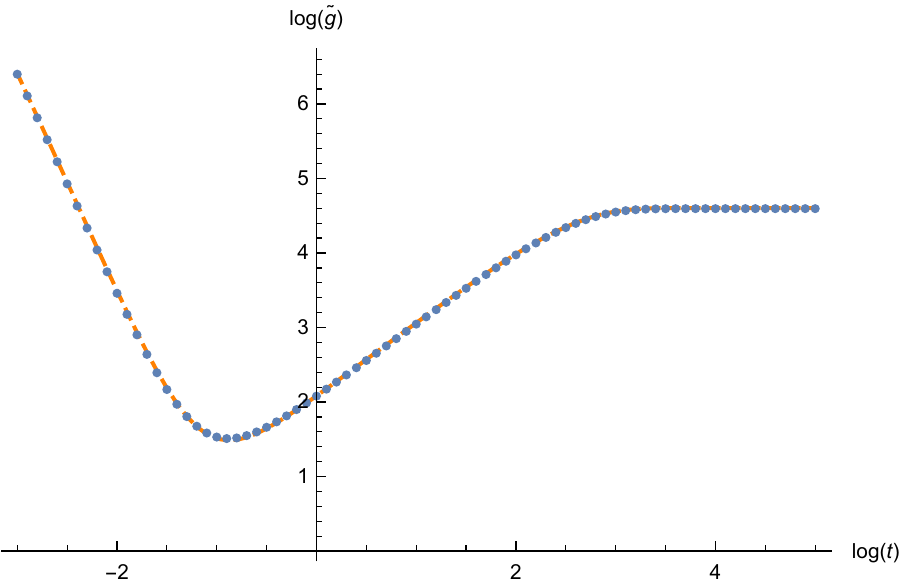}}
\hskip5mm
\subcaptionbox{$E'=3$ \label{sfig:sff-e3}}{\includegraphics
[width=0.3\linewidth]{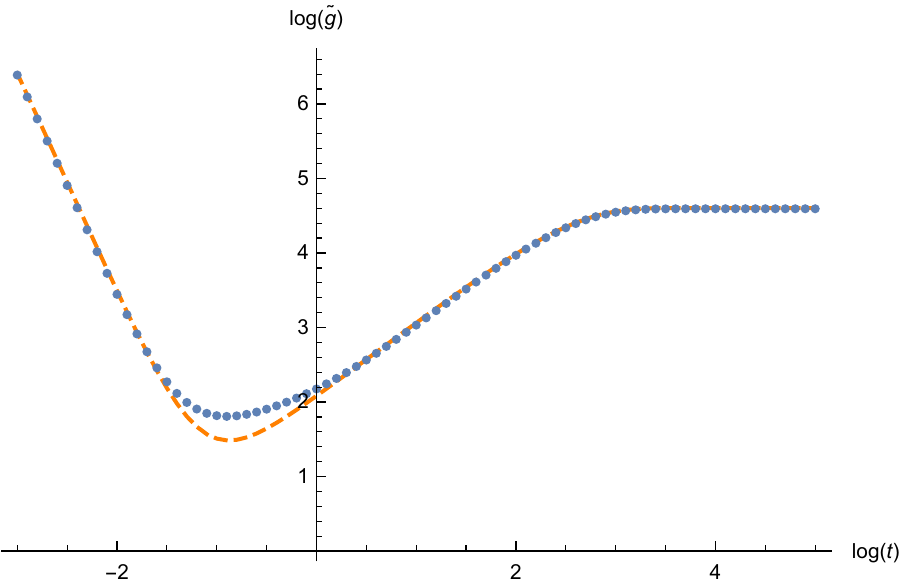}}
\subcaptionbox{$E'=20$ \label{sfig:sff-e20}}{\includegraphics
[width=0.3\linewidth]{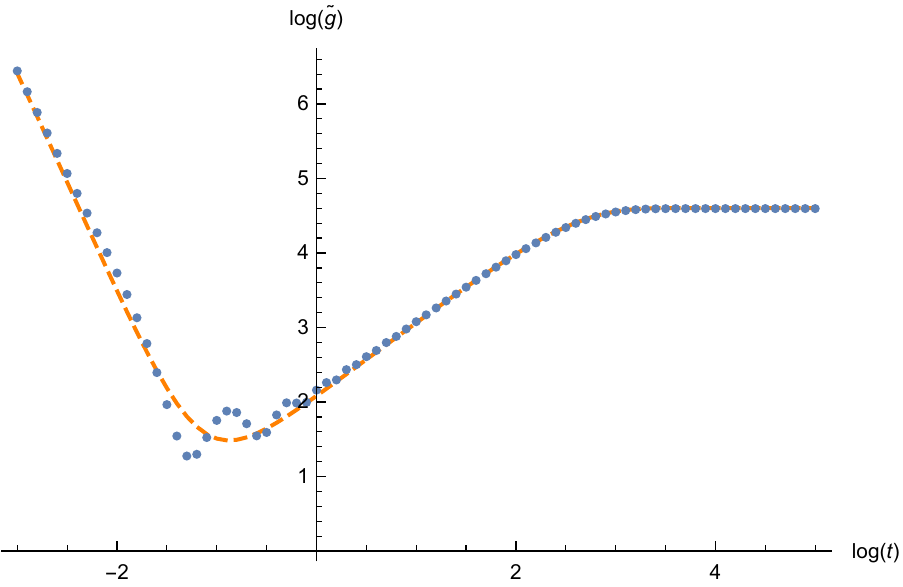}}
  \caption{
Plot 
of SFF $\til{g}(t,\bt,E')$ with $\bt=1/100$
for 
\subref{sfig:sff-em3} $E'=-3$,  \subref{sfig:sff-e3} $E'=3$ and
\subref{sfig:sff-e20} $E'=20$
as a function of $t$.
The blue dots represent the SFF in the presence of two FZZT branes
at $\xi=-E'$,
while
the orange dashed curve represents the SFF without the FZZT branes.
}
  \label{fig:SFF}
\end{figure}

Let us consider the two-point function of macroscopic loops in the presence of two FZZT
branes
\begin{equation}
\begin{aligned}
 g(\bt_1,\bt_2,\xi)&= \frac{\bra Z(\bt_1)Z(\bt_2)\Psi(\xi)^2\ket}{\bra \Psi(\xi)^2\ket}.
\end{aligned} 
\end{equation}
As in the previous subsection, we put two FZZT branes since
single FZZT brane is not positive definite.
From our general formula \eqref{eq:even-gen} we find
\begin{equation}
\begin{aligned}
 g(\bt_1,\bt_2,\xi)
&=\bra Z(\bt_1)Z(\bt_2)\ket\\
&+\frac{1}{K(\xi,\xi)}\Biggl[-\bra \xi|\Pi e^{(\bt_1+\bt_2)Q}\Pi|\xi\ket
+2\bra \xi|\Pi e^{\bt_1Q}\Pi e^{\bt_2Q}\Pi|\xi\ket\\
&\qquad\qquad-\bra Z(\bt_1)\ket \bra \xi|\Pi e^{\bt_2Q}\Pi|\xi\ket-
\bra Z(\bt_2)\ket \bra \xi|\Pi e^{\bt_1Q}\Pi|\xi\ket\Biggr].
\end{aligned} 
\end{equation}
We can show that this is rewritten as
\begin{equation}
\begin{aligned}
g(\bt_1,\bt_2,\xi)
&=\int d\la e^{(\bt_1+\bt_2)\la}\frac{1}{K(\xi,\xi)}
\left|
\begin{matrix}
K(\la,\la) & K(\la,\xi)\\
K(\xi,\la) &K(\xi,\xi) 
\end{matrix}
\right|\\
&+\int\prod_{i=1}^2 d\la_i e^{\bt_i\la_i}
\frac{1}{K(\xi,\xi)}
\left|
\begin{matrix}
K(\la_1,\la_1) & K(\la_1,\la_2)&K(\la_1,\xi)\\
K(\la_2,\la_1) &K(\la_2,\la_2)&K(\la_2,\xi)\\
 K(\xi,\la_1) &K(\xi,\la_2)&K(\xi,\xi)
\end{matrix}
\right|.
\end{aligned}
\label{eq:SFF-integral}
\end{equation}
We consider the spectral form factor (SFF) in the presence of two FZZT
branes $\Psi(\xi)^2$ at $\xi=-E'$
\begin{equation}
\begin{aligned}
 \til{g}(t,\bt,E')=g(\bt+\ri t,\bt-\ri t,-E').
\end{aligned} 
\label{eq:SFF}
\end{equation}
Using the CD kernel for the Airy case \eqref{eq:CD-Airy}
and the integral representation \eqref{eq:SFF-integral}, 
we can evaluate this SFF \eqref{eq:SFF} numerically.
In Figure~\ref{fig:SFF} we show the plot of SFF
in the Airy case with two FZZT branes for $\bt=1/100$
with several different values of $E'$.
For $E'=-3$ and $E'=3$, we do not see a notable
deviation from the original
SFF without the FZZT branes (orange dashed curve in Figure~\ref{fig:SFF}),
but the $E'=20$ case in Figure
\ref{sfig:sff-e20} has some oscillatory behavior. 
As discussed in \cite{Blommaert:2019wfy},
we expect that the erratic behavior arises if we put many FZZT branes.
It is tempting to speculate that the oscillatory behavior in Figure
\ref{sfig:sff-e20} might be an indication towards the erratic behavior
as we increase the number of FZZT branes.
It would be interesting to study the SFF with many FZZT brane insertions
explicitly in the Airy case. 
However, this is a challenging problem since the generalization
of \eqref{eq:SFF-integral} for many FZZT brane insertions is a multi-variable integral
which is not easy to evaluate numerically with high precision.
We leave this as an interesting future
problem.

\section{Conclusions and outlook}\label{sec:conclusion}
In this paper we have studied FZZT branes in JT gravity
and topological gravity.
We found that FZZT branes can be introduced in the matrix model of JT gravity
\cite{Saad:2019lba} 
by attaching $\cM(b)=-e^{-zb}$ to the geodesic boundary
of length $b$ and integrating over $b$ (see Figure~\ref{fig:trumpet-FZZT}).
Our construction can be generalized to arbitrary background $\{t_k\}$
of topological gravity by introducing the generalized WP volume
\eqref{eq:Vgn-operator}. We argued that the trumpet ending on a FZZT brane
can be thought of as a ``half-wormhole'' introduced in \cite{Saad:2021rcu}
(see Figure~\ref{fig:half-wormhole}).
We found that the half-wormhole amplitude is given by the complementary error function
\eqref{eq:z-HW}.
FZZT branes induce the shift of couplings \eqref{eq:tilde-tk}
which is consistent with the known property of BA function 
in the literature.
However, the expression such as \eqref{eq:BA-tau} only gives an asymptotic expansion
of the BA function in a certain sector of the complex $z$-plane
\eqref{eq:arg-z}. In section~\ref{sec:corr} we found the general
formulae \eqref{eq:even-gen}
and \eqref{eq:odd-gen}
for the generating function of 
the correlator of FZZT branes and macroscopic loops.
Our formula expresses the correlator of FZZT branes and macroscopic loops
in terms of the BA function and the CD kernel, which in principle can 
be defined non-perturbatively.
As an example, in section~\ref{sec:airy} we studied the 
eigenvalue density and the spectral form factor (SFF) deformed by 
the insertion of two FZZT branes in the Airy case.
In this explicit computation, we 
confirmed the picture of ``eigenbranes'' put forward in \cite{Blommaert:2019wfy}.

There are many interesting open questions. It would be interesting to
generalize our study of SFF by inserting more FZZT branes to see if we 
recover the erratic behavior of SFF
as anticipated in \cite{Blommaert:2019wfy}. 
Another interesting direction is the derivation of the Page curve 
in our formalism of FZZT branes.
In \cite{Penington:2019kki}, it is argued that the Page curve is
reproduced by adding EOW brane degrees of freedom to JT gravity.
As we have seen in \eqref{eq:EOW}, the EOW brane in \cite{Gao:2021uro} is a 
special case of our formalism, corresponding to an infinite collection of
anti-FZZT branes.
It would be interesting to apply our formalism of FZZT branes to this problem.
We should also mention that as argued in \cite{Maldacena:2004sn}
the ``meson'' field $S$ constructed out of the fermions in \eqref{eq:det-fermion}
\begin{equation}
\begin{aligned}
 S_{ij}=\b{\chi}_{ia}\chi_{aj}
\end{aligned} 
\label{eq:Sij}
\end{equation}
can be identified as the matrix appearing in the Kontsevich model
\begin{equation}
\begin{aligned}
 e^{F(\bm{t})}&=\int dS e^{\frac{\ri}{2\gs}\Tr \left(\frac{S^3}{3}+ZS^2\right)},\\
t_k&=-\gs(2k-1)!!\Tr Z^{-2k-1}.
\end{aligned} 
\label{eq:Kon}
\end{equation} 
It would be interesting to study, say, the second R\'{e}nyi entropy
\begin{equation}
\begin{aligned}
 e^{-S_2}=\frac{\bra\Tr S^2\ket}{\bra\Tr S\ket^2}
\end{aligned} 
\end{equation} 
in Kontsevich model in the JT gravity background with FZZT branes
\eqref{eq:JT-FZZT} to see if the Page curve is recovered.
We leave this as an interesting future problem.

\acknowledgments
We would like to thank Douglas Stanford for correspondence.
This work was supported in part by JSPS KAKENHI Grant
Nos.~19K03845 and 19K03856,
and JSPS Japan-Russia Research Cooperative Program.

\appendix
\section{\mathversion{bold}$(2,p)$ minimal model background}\label{app:minimal}
The background for the $(2,p)$ minimal string is given by
\cite{Mertens:2020hbs,Turiaci:2020fjj}
\begin{equation}
\begin{aligned}
 t_k-\cob_{k,1}&=\frac{(-1)^k(m+k-2)!}{(k-1)!(m-k)!}\left(\frac{2}{2m-1}\right)^{2k-2},
\end{aligned} 
\label{eq:tk-min}
\end{equation}
where $p$ and $m$ are related by $p=2m-1$.
One can see that $t_k$ in \eqref{eq:tk-min} reduces to $\ga_k$ in \eqref{eq:ga-JT}
in the limit 
$m\to\infty$.
In this sense, JT gravity is a $p\to\infty$ limit of the $(2,p)$
minimal string theory.

The Itzykson-Zuber variable $I_0$ for the minimal model background
is
\begin{equation}
\begin{aligned}
 I_0(u)-u&=\sum_{k=1}^m(t_k-\cob_{k,1})\frac{u^k}{k!}\\
&=\frac{p}{8}\left[P_m\left(1-\frac{8u}{p^2}\right)-P_{m-2}\left(1-\frac{8u}{p^2}\right)\right],
\end{aligned} 
\end{equation}
where $P_m(x)$ denotes the Legendre polynomial.

The genus-zero eigenvalue density is given by
\begin{equation}
\begin{aligned}
 \rho_0(E)&=\frac{1}{\pi \gs}\sum_{k=1}^m\frac{(-1)^k(t_k-\cob_{k,1})}{(2k-1)!!}(2E)^{k-\hf}\\
&=\frac{1}{\rt{2}\pi \gs}\sinh\left[p\,\text{arcsinh}\left(\frac{2\rt{E}}{p}\right) 
\right],
\end{aligned} 
\label{eq:rho-min}
\end{equation}
and the disk amplitude is given by
\begin{equation}
\begin{aligned}
 \bra Z(\bt)\ket^{g=0}&=\frac{1}{\rt{2\pi}\gs}\sum_{k=1}^m (-1)^k(t_k-\cob_{k,1})\bt^{-k-\hf}\\
&=\frac{1}{\rt{2}\pi \gs}\frac{p}{2\bt}e^{\frac{\bt p^2}{8}}K_{\frac{p}{2}}\Bigl(\frac{\bt p^2}{8}\Bigr).
\end{aligned} 
\end{equation}
One can easily see that in the $p\to\infty$ limit the eigenvalue density
of the $(2,p)$ minimal string \eqref{eq:rho-min} reduces
to the eigenvalue density of JT gravity in \eqref{eq:JT-rho}.

\section{\mathversion{bold}$Z(\bt)$-FZZT amplitude from inverse determinant}\label{app:inverse}
We can reproduce the $Z(\bt)$-FZZT amplitude
\eqref{eq:Z-FZZT-full} using the relation
\begin{equation}
\begin{aligned}
 \left\bra\Tr\frac{1}{x-M}\det(\xi-M)\right\ket=\lim_{y\to x}\del_y
\left\bra\frac{\det(y-M)\det(\xi-M)}{\det(x-M)}\right\ket.
\end{aligned} 
\end{equation}
The correlator of determinants and inverse determinants
is studied in \cite{Fyodorov:2002jw,Strahov:2002zu,Baik_2003}.
Using the result of \cite{Fyodorov:2002jw,Strahov:2002zu,Baik_2003}, we find
\begin{equation}
\begin{aligned}
 \left\bra\frac{\det(y-M)\det(\xi-M)}{\det(x-M)}\right\ket=
\frac{1}{\xi-y}
\frac{1}{h_{N-1}}\left|
\begin{matrix}
 \varphi_{N-1}(x)&\varphi_{N}(x)&\varphi_{N+1}(x)\\
P_{N-1}(y) & P_N(y) &P_{N+1}(y)\\
P_{N-1}(\xi) & P_N(\xi) &P_{N+1}(\xi)
\end{matrix}
\right|,
\end{aligned} 
\end{equation}
where $\varphi_N(x)$ is the Hilbert transform of the orthogonal polynomial
$P_N(x)$
\begin{equation}
\begin{aligned}
 \varphi_N(x)=\int d\la \frac{e^{-V(\la)}}{x-\la}P_N(\la).
\end{aligned} 
\end{equation}
Thus we find
\begin{equation}
\begin{aligned}
 \left\bra\Tr\frac{1}{x-M}\det(\xi-M)\right\ket=
\frac{1}{\xi-x}
\frac{1}{h_{N-1}}\int d\la \frac{e^{-V(\la)}}{x-\la}
\left|
\begin{matrix}
 P_{N-1}(\la)& P_{N}(\la)& P_{N+1}(\la)\\
P_{N-1}'(x) & P_N'(x) &P_{N+1}'(x)\\
P_{N-1}(\xi) & P_N(\xi) &P_{N+1}(\xi)
\end{matrix}
\right|+\frac{P_N(\xi)}{\xi-x}.
\end{aligned}
\label{eq:dd/d-delta} 
\end{equation}
By using
the relation
\begin{equation}
\begin{aligned}
 \la P_N(\la)=P_{N+1}(\la)+\frac{h_N}{h_{N-1}}P_{N-1}(\la),
\end{aligned} 
\end{equation}
the
first term of \eqref{eq:dd/d-delta} is written as
\begin{equation}
\begin{aligned}
 &\frac{1}{\xi-x}\frac{1}{h_{N-1}}\int d\la \frac{e^{-V(\la)}}{x-\la}
\left|
\begin{matrix}
 P_{N-1}(\la)& P_{N}(\la)& \la P_{N}(\la)\\
P_{N-1}'(x) & P_N'(x) & xP_{N}'(x)+P_N(x)\\
P_{N-1}(\xi) & P_N(\xi) &\xi P_{N}(\xi)
\end{matrix}
\right|\\
&=\frac{1}{\xi-x}\frac{1}{h_{N-1}}\int d\la \frac{e^{-V(\la)}}{x-\la}
\left|
\begin{matrix}
 P_{N-1}(\la)& P_{N}(\la)& (\la-x) P_{N}(\la)\\
P_{N-1}'(x) & P_N'(x) & P_N(x)\\
P_{N-1}(\xi) & P_N(\xi) &(\xi-x) P_{N}(\xi)
\end{matrix}
\right|\\
&=\frac{1}{h_{N-1}}\int d\la \frac{e^{-V(\la)}}{x-\la}
\Biggl[P_N(\xi)\Bigl(P_{N-1}(\la)P_N'(x)-P_N(\la)P_{N-1}'(x)\Bigr)\\
&\quad -P_N(x)\frac{P_{N-1}(\la)P_N(\xi)-P_{N}(\la)P_{N-1}(\xi)}{\xi-x}\Biggr].
\end{aligned} 
\label{eq:d/d-P}
\end{equation}
In the last equality we used the orthogonality relation \eqref{eq:Pn-orth}.
Let us consider the first half
of \eqref{eq:d/d-P}
\begin{equation}
\begin{aligned}
 &\frac{1}{h_{N-1}}\int d\la \frac{e^{-V(\la)}}{x-\la}
P_N(\xi)\Bigl(P_{N-1}(\la)P_N'(x)-P_N(\la)P_{N-1}'(x)\Bigr)\\
&=\frac{1}{h_{N-1}}\int d\la \frac{e^{-V(\la)}}{x-\la}
P_N(\xi)\Bigl(P_{N-1}(\la)P_N'(\la)-P_N(\la)P_{N-1}'(\la)\Bigr)
\\
&\quad+\frac{P_N(\xi)}{h_{N-1}}\int d\la e^{-V(\la)}\Biggl[P_{N-1}(\la)\frac{P_N'(x)-P_N'(\la)}{x-\la}
-P_N(\la)\frac{P_{N-1}'(x)-P_{N-1}'(\la)}{x-\la}\Biggr] \\
&=\frac{1}{h_{N-1}}\int d\la \frac{e^{-V(\la)}}{x-\la}
P_N(\xi)\Bigl(P_{N-1}(\la)P_N'(\la)-P_N(\la)P_{N-1}'(\la)\Bigr).
\end{aligned} 
\end{equation}
In the last step we used the orthogonality relation \eqref{eq:Pn-orth}.
The second half
of \eqref{eq:d/d-P} is
\begin{equation}
\begin{aligned}
& \frac{1}{h_{N-1}}\int d\la \frac{e^{-V(\la)}}{x-\la}P_N(x)\frac{P_{N-1}(\la)P_N(\xi)-P_{N}(\la)P_{N-1}(\xi)}{x-\xi}\\
=&\frac{1}{h_{N-1}}\int d\la e^{-V(\la)}\frac{P_N(x)-P_N(\la)+P_N(\la)}{x-\la}
\frac{P_{N-1}(\la)P_N(\xi)-P_{N}(\la)P_{N-1}(\xi)}{x-\xi}.
\end{aligned} 
\label{eq:2ndterm}
\end{equation}
Using the relation
\begin{equation}
\begin{aligned}
 \frac{P_N(x)-P_N(\la)}{x-\la}=P_{N-1}(\la)+(\text{lower degree terms in}~ \la)
\end{aligned} 
\end{equation}
\eqref{eq:2ndterm} becomes
\begin{equation}
\begin{aligned}
 &\frac{P_N(\xi)}{x-\xi}+
\frac{1}{h_{N-1}}\int d\la e^{-V(\la)}\frac{P_N(\la)}{x-\la}
\frac{P_{N-1}(\la)P_N(\xi)-P_{N}(\la)P_{N-1}(\xi)}{x-\xi}\\
=&\frac{P_N(\xi)}{x-\xi}+\frac{1}{h_{N-1}}\int d\la
e^{-V(\la)}\left(\frac{1}{x-\xi}-\frac{1}{x-\la}\right)
P_N(\la)\frac{P_{N-1}(\la)P_N(\xi)-P_{N}(\la)P_{N-1}(\xi)}{\xi-\la}\\
=&\frac{P_N(\xi)}{x-\xi}-\frac{1}{h_{N-1}}\int d\la e^{-V(\la)}\frac{P_N(\la)}{x-\la}
\frac{P_{N-1}(\la)P_N(\xi)-P_{N}(\la)P_{N-1}(\xi)}{\xi-\la}.
\end{aligned} 
\end{equation}
The first term cancels out the last term of \eqref{eq:dd/d-delta}.
Finally we find
\begin{equation}
\begin{aligned}
 \left\bra\Tr\frac{1}{x-M}\det(\xi-M)\right\ket=
\frac{1}{h_{N-1}}\int & d\la \frac{e^{-V(\la)}}{x-\la}
\Biggl[P_N(\xi)\Bigl(P_N'(\la)P_{N-1}(\la)-P_{N}(\la)P_{N-1}'(\la)\Bigr)\\
&-P_N(\la)\frac{P_N(\xi)P_{N-1}(\la)-P_{N-1}(\xi)P_{N}(\la)}{\xi-\la}\Biggr].
\end{aligned} 
\end{equation}
In the double scaling limit this becomes
\begin{equation}
\begin{aligned}
 \left\bra\Tr\frac{1}{x-M}\Psi(\xi)\right\ket=
\int \frac{d\la}{x-\la}\Bigl[K(\la,\la)\psi(\xi)-K(\xi,\la)\psi(\la)\Bigr].
\end{aligned} 
\end{equation}
This reproduces the deformed eigenvalue density due to the insertion of one FZZT brane
in \eqref{eq:Z-FZZT-full}.

One can in principle derive the multi-point correlator of $Z(\bt)$'s and FZZT branes
in this approach using the result in \cite{Fyodorov:2002jw,Strahov:2002zu,Baik_2003}.
We leave this as an interesting future problem.

\bibliography{paper}
\bibliographystyle{utphys}

\end{document}